\documentclass{article}
\usepackage{amsfonts}
\usepackage{amsmath}
\usepackage{amssymb}

\input{tcilatex}
\begin{document}

\title{Lorentz group in gravity theories}
\author{Jianbo Lu$^{1\ast }$, Yongxin Guo$^{2\dag }$ and G. Y. Chee$^{1\ddag
}$ \and $^{1}$Department of Physics, Liaoning Normal
University, Dalian, China \and $^{2}$Department of Physics, Liaoning
University, Shenyang, China \and $^{\ast }$lvjianbo819@163.com \\
$^{\dag }$guoyongxin@lnu.edu.cn \and $^{\ddag }$qgy8475@sina.com}
\maketitle

\begin{abstract}
In this paper, it is argued that in gravity theories the local Lorentz group
can not be considered as a gauge group in the sense of Yang-Mills theories,
the Lorentz connection is not a gauge potential but an artificial force, the
inertial force. A genuine gravity theory should be a translation gauge
theory, though a unnormal gauge theory. All the three theories of the
Geometrical Trinity of Gravity are translation gauge theories. A real
gravity theory should get rid of "gauging" Lorentz group. The
covariantization of the teleparallel gravity is not necessary physically.
\end{abstract}

\section{Introduction}

In recent years the\textbf{\ alternative} formulations of General Relativity
(GR) [1] in terms of torsion and non-metricity have received considerable
attention. They not only provide novel and fresh perspectives at various
aspects of gravity, but also open new avenues to extend GR. Although the
three\textsl{\ }\textquotedblleft pictures\textquotedblright\ of GR among
the geometrical trinity of gravity are equivalent, the modified gravity
theories developed from them, such as $f(R)$, $f(T)$, $f(Q)$ theories, are
very different from each other.\textsl{\ }And despite a large number of
reasonable ideas on these theories, we cannot claim a definitive success of
any particular approach in them.\ The search for a viable modification to
the theory of gravitational interactions is anything but a simple task. Even
a lot of modified gravity theories fail already at the purely theoretical
level before considering their potential phenomenological applications. The
rapid progress of experiment and theory has led to profound changes in many
ideas, but it has also created new confusions. New disagreements arose over
many basic concepts. For example, what is gravity? According to the
traditional view, "gravity is the curvature of the spacetime". However, the
teleprallel equivalent of general relativity (TEGR) [2,3] indicates that
gravity can be described by either a curved Riemann spacetime or a flat
Witzenb\"{o}ck spacetime with the torsion. Furthermore, the symmetric
teleparallel equivalent of general relativity (STEGR) [4] or the coincident
general relativity (CGR) [5] demonstrates that the description of gravity
needs neither the curvature nor the torsion but only the nonmetricity.

TEGR and STEGR overthrow the \textbf{dogma}, "gravity is the curvature of
the spacetime". The fundamental variable describing gravity is the metric or
the tetrad, which represent the gravitational potential. The gravitational
field strength is their first derivatives, the non-metricity or the torsion
rather than the curvature, since in general case curvature has nothing to do
with either metric or tetrad [6]. Therefore, it is asserted that geometrical
description of gravity\ is merely a matter of convention. There are
equivalent formulations in terms of different fields in different
geometrical frameworks [7, 8, 9,10].

However, what is gravity? The question remains unanswered. It is generally
accepted that gravity is a gauge field. However, what is the gauge group of
the gravitational interaction? It is still far from consensus. An
authoritative view is that gravity is a translation gauge field [11], but it
can only be really understood in the context of Poincare gauge. The
translation gauge approach is not comprehensible as a stand-alone theory
[12].

Usually, the teleprallel equivalent of general relativity is considered as a
translation gauge theory [2]. Recently, however, it is argued that gravity
theory as a translation gauge theory can not be formulated with precisely
the structure of a Yang-Mills type [13]. The translations is gauged in
Lorentz gauge theory completely different. In contrast to conventional gauge
theories of gravity [14], the Lorentz gauge theory is formulated based on a
different approach akin to parameterised field theories [15].

One of the very hot topics is about the issue of local Lorentz invariance in
modified teleparallel gravities. There are many discussions on the covariant
approach [16, 17] to these theories. The conclusions range all the way from
total denial [18] through an undecided stance [19] to taking it as the only
consistent version of the theory [20]. The recent discussion see [21,22],
for example.

Various kinds of arguments and disagreements in gravitational gauge theories
focus on gauging of the Lorentz group. Since 1956 Utiyama [23] introduced
gauged Lorentz group into gravitational theories the controversy about its
role has continued to this day. The key point is that different from other
gauge transformations, Lorentz transformation includes time in itself, and
therefor includes the dynamic effects of the reference frame. The Lorentz
connection possesses different physical meaning from other connections. \
The Lorentz connection represents a force (inertial force), while other
connections represent a potential.

Lorentz symmetry is a cornerstone of modern physics. The Standard Model is
formulated as a quantum field theory based on the global Lorentz symmetry of
Special Relativity. The\ fields is classified according to the
representations of the Lorentz group [24]. However, in all of these
Yang-Mills type gauge theories of particles the Lorentz group has never been
gauged. In fact, in a relativistic gravity theory, a global Lorentz\textsl{\ 
}group is enough, the gauging of it is not necessary. Not only that, but it
is the source of almost all the trouble.

The real sense of TEGR and CGR is getting rid of the local Lorentz group or
affine group, these two theories respectively\ propose a simpler geometrical
formulation of GR that is oblivious to the affine spacetime structure, thus
fundamentally depriving gravity of any inertial character [5]. It is
asserted that the affine connection is the recognized as a\ purely
fictitious force [25-28]. Especially, except in general relativity, in\ all
other relativistic theories the Lorentz connection\ has to do with inertial
effects only [29]. In other words, the affine connection, including the
Lorentz connection as a special case represents only a\ fictitious force,
i.e. an inertial force rather than a genuine gravitational one.

Decades of discussion have brought to light many significant points, which
we intend to bring together into a unified view. This article will
demonstrate this point in detail. With this consideration, a series of
fundamental concepts and principles in geometry and gravity will be
discussed. In Sec. 2, the geometric tools for the formulation of gravity
theories are reviewed. Especially, the affine connection and the translation
connection are introduced. The affine connection has two patterns of
manifestation, i.e. the frame ("internal") patterns $\Gamma ^{I{}}{}_{J\mu }$
and the coordinate ("external") patterns ("internal") $\Gamma ^{\lambda
}{}_{\mu \nu }$. The relation between them is clarified. The \textbf{torsion
of a frame} is defined. The relation between torsion and curvature is
clarified. Some conceptual remarks are proposed. In Sec. 3, we argue that
either the\textbf{\ }\textsl{local} Lorentz group $SO(1,3)$ or the\ local
general linear group $GL(4,R)$ can never be considered as a gauge group,
since their connection does not represent a potential of some force, but a
force itself, the inertial force. The genuine gravitational gauge group is
the translation group $T^{4}$. The diffeomorphism or the Lorentz group
implements the relativity principle, whereas the translation group
implements the gauge principle and introduces the gravitational interaction.
As a translation gauge theory, gravity theories are classified as two kinds
of formulations, the metric formulation and the tetrad formulation, instead
of the geometry trinity. In Subsec. 3.3, a concise historical review is
given to show how the various theories of gravity converge to the
translation gauge theory. In Sec. 4 some traditional ideas are reexamined.
Sec. 5 is devoted to conclusions.

\section{Geometry}

Let $M$ be a 4-dimensional differentiable manifold with a tangent space $%
T_{x}M$ at each point $x\in M$. Consider a coordinate system $\{x^{\mu }\}$
on this manifold, and also a coordinate system $\{y^{I}\}$ on the tangent
space. Such coordinate systems define, on their domains of definition, local
bases for vector fields formed by the sets of gradients $\{\frac{\partial }{%
\partial x^{\mu }}\}$, $\{\frac{\partial }{\partial y^{I}}\}$, as well as
bases $\{dx^{\mu }\}$, $\{dy^{I}\}$ for covector fields or differential
forms. The tangent space $T_{x}M$ is a fiber of the\ tangent bundle with the
the base manifold $M$ and the structure group $GL\left( 4,\mathbb{R}\right) $%
.

Any set of four linearly independent vector fields $\{b_{I}\}$ on $T_{x}M$
will form a base, and will have a dual $\{b^{I}\}$ satisfying $%
b^{I}(b_{J})=\delta _{J}^{I}$. These frame fields are the general linear
bases on $T_{x}M$ whose set is also a differentiable manifold, and
constitutes the bundle of linear frames, the principal bundle\ associated
with the tangent bundle. On the common domains they are defined, the members
of a base $\{b_{I}\}$ can be written in terms of the members of the other $%
\{\partial _{%
\mu
}=\frac{\partial }{\partial x^{\mu }}\}$: $b_{I}=b_{I}{}^{%
\mu
}\partial _{%
\mu
}$, $b^{I}=b^{I}{}_{%
\mu
}dx^{%
\mu
}$, and conversely, $\partial _{%
\mu
}=b^{I}{}_{%
\mu
}b_{I}$, $dx^{%
\mu
}=b_{I}{}^{%
\mu
}b^{I}$. Any vector $V\in T_{x}M$ can be expressed as $V=V^{I}b_{I}\ $as
well as $V=V^{\mu }\partial _{\mu }$. Notice that these frames, with their
bundle, are\ constitutive parts of the differentiable manifold [6, 30]. The
whole bundle space is locally the Cartesian product $M\times G$, where the
structure group $G$ of the bundle is the general linear group $GL(4,R)$.
Once manifold $M$ is endowed with a Lorentz metric $g$, one can define a
sub-bundle called the bundle of orthonormal frames, which consists of
orthonormal bases with respect to $g$, and with the Lorentz group $SO(1,3)$
as the structure group [31]. Technically, the orthonormality condition for
the frames is introduced with the help of the soldering form [32].

\subsection{Affine geometry}

\subsubsection{Local general linear group $G(4,\mathbb{R})$, affine
connection}

Let $\left\{ b_{K}\right\} $ and $\left\{ b_{K}^{\prime }\right\} $ are two
basis (\textbf{frames}) of $T_{x}M$. Any vector $V\in T_{x}M$ can be
expressed as%
\begin{equation}
V=V^{K}b_{K}=V^{\prime K}b_{K}^{\prime }.
\end{equation}%
Under the base transformation

\begin{equation}
b_{K}^{\prime }=\Lambda _{K}{}^{L}b_{L},\ \ \ \Lambda _{K}{}^{L}\in GL(4,%
\mathbb{R})\ 
\end{equation}%
the components $V^{K}$ of a vector transform as%
\begin{equation}
V^{\prime K}=V^{L}\left( \Lambda ^{-1}\right) _{L}{}^{K},
\end{equation}%
where $\Lambda ^{-1}$ is the inverse of the matrix $\Lambda $. Therefore,
the vector $V=V^{K}b_{K}$ is usually called a cotravariant vector. Vectors $%
V^{K}$ defined at every points on $M$ form a field, which means that $%
V^{K}=V^{K}\left( x\right) $ and $\Lambda _{K}{}^{L}=\Lambda
_{K}{}^{L}\left( x\right) $,\ as well as $b_{K}=b_{K}\left( x\right) $ and $%
b_{K}^{\prime }=b_{K}^{\prime }\left( x\right) $ are functions of the
coordinates $x^{\mu }$ of $x$. Under the transformation $\left( 1\right) $,
the derivative transforms also:%
\begin{equation*}
\partial _{\mu }V^{\prime K}=\partial _{\mu }V^{L}\left( \Lambda
^{-1}\right) _{L}{}^{K\prime }+V^{L}\partial _{\mu }\left( \Lambda
^{-1}\right) _{L}{}^{K\prime }.
\end{equation*}%
It is not covariant with $\left( 2\right) $ due to the second term. However,
if there is an affine connection $\Gamma ^{K}{}_{L\mu }=\Gamma ^{K}{}_{L\mu
}\left( x\right) $ defining the covariant derivative\textsl{\ } 
\begin{equation}
D_{\mu }V^{K}:=\partial _{\mu }V^{K}+\Gamma ^{K}{}_{L\mu }V^{L}.\ \ \ 
\end{equation}%
and obeying the transformation rules%
\begin{equation}
\Gamma ^{\prime K}{}_{I\mu }=\left( \Lambda ^{-1}\right) _{L}{}^{K}\Gamma
^{L}{}_{J\mu }\Lambda _{I}{}^{J}+\left( \Lambda ^{-1}\right)
_{J}{}^{K}\partial _{\mu }\Lambda _{I}{}^{J},
\end{equation}%
then%
\begin{equation}
D_{\mu }V^{\prime K}=D_{\mu }V^{L}\left( \Lambda ^{-1}\right) _{L}{}^{K}.\ \
\ \ \ \ \ \ \ \ \ \ \ \ 
\end{equation}%
is covariant with ($3$), which is the reason for name of the "covariant"
derivative $D_{\mu }$.\ 

The frame bundle is associated to the tangent bundle, $\Gamma ^{K}{}_{L\mu }$
is the connection of the frame bundle as well as of the tangent bundle.

\textbf{Two examples:}

\textsl{i)} Under a coordinate transformation $x^{\mu }\rightarrow x^{\prime
\mu }$, the coordinate basis transforms as%
\begin{equation*}
\partial _{\mu }\rightarrow \partial _{\mu }^{\prime }=\frac{\partial x^{\nu
}}{\partial x^{\prime \mu }}\partial _{\nu }.
\end{equation*}%
By the substitutions%
\begin{equation*}
\ \Lambda _{K}{}^{L}\Longrightarrow \frac{\partial x^{\tau }}{\partial
x^{\prime \lambda }},\left( \Lambda ^{-1}\right) _{J}{}^{K}\Longrightarrow 
\frac{\partial x^{\prime \lambda }}{\partial x^{\tau }},\Gamma ^{L}{}_{J\mu
}\Longrightarrow \Gamma ^{\rho }{}_{\nu \mu },
\end{equation*}%
$\left( 5\right) $ takes the form%
\begin{equation}
\Gamma ^{\prime \lambda }{}_{\nu \mu }=\frac{\partial x^{\prime \lambda }}{%
\partial x^{\rho }}\frac{\partial x^{\sigma }}{\partial x^{\prime \nu }}%
\frac{\partial x^{\tau }}{\partial x^{\prime \mu }}\Gamma ^{\rho }{}_{\sigma
\tau }+\frac{\partial x^{\prime \lambda }}{\partial x^{\tau }}\frac{\partial
^{2}x^{\tau }}{\partial x^{\prime \nu }\partial x^{\prime \mu }}.
\end{equation}

In this case the covariant derivative $\left( 4\right) $ takes the form%
\begin{equation}
\nabla _{\mu }V^{\alpha }=\partial _{\mu }V^{\alpha }{}+\Gamma ^{\alpha
}{}_{\mu \lambda }V^{\lambda },\;\;\;\;\;\;\;\;\;
\end{equation}%
for a vector $V^{\alpha }$ and 
\begin{equation}
\;\nabla _{\mu }V_{\alpha }=\partial _{\mu }V{}_{\alpha }-\Gamma ^{\lambda
}{}_{\mu \alpha }V_{\lambda }.\;\;\;\;\;\;\;\;
\end{equation}%
for a 1-forms (covector) $V_{\alpha }$.

\textsl{ii)} Considering the decompsition%
\begin{equation*}
b_{I}=b_{I}{}^{\mu }\partial _{\mu }
\end{equation*}%
as a transformation between $\left\{ \partial _{\mu }\right\} $ and $\left\{
b_{I}\right\} $, by the substitutions 
\begin{equation*}
b_{K}^{\prime }\Longrightarrow \partial _{\mu },\Gamma ^{\prime K}{}_{I\mu
}\Longrightarrow \Gamma ^{\rho }{}_{\mu \nu }
\end{equation*}%
$(4)$ and $(5)$ can be written as%
\begin{equation}
D_{\mu }b^{I}{}_{\nu }=\partial _{\mu }b^{I}{}_{\nu }+\Gamma ^{I}{}_{J\mu
}b^{J}{}_{\nu },
\end{equation}%
and%
\begin{equation}
\Gamma ^{\rho }{}_{\mu \nu }=b_{I}{}^{\rho }\partial _{\mu }b^{I}{}_{\nu
}+b_{I}{}^{\rho }\Gamma ^{I}{}_{J\mu }b^{J}{}_{\nu }\equiv b_{I}{}^{\rho
}D_{\mu }b^{I}{}{}_{\nu }.
\end{equation}%
The inverse relation is, consequently, 
\begin{equation}
\Gamma ^{I}{}_{J\mu }=b^{I}{}_{\nu }\partial _{\mu }b_{J}{}^{\nu
}+b^{I}{}_{\rho }\Gamma ^{\rho }{}_{\mu \nu }b_{J}{}^{\nu }\equiv
b^{I}{}_{\nu }\nabla _{\mu }b_{J}{}^{\nu },
\end{equation}%
with%
\begin{equation}
\nabla _{\mu }b_{I}{}^{\nu }=\partial _{\mu }b_{I}{}^{\nu }+\Gamma ^{\nu
}{}_{\mu \rho }b_{I}{}^{\rho }.
\end{equation}

\subsection{A special class of tensors}

Under the base transformation

\begin{equation*}
b_{K}^{\prime }=\Lambda _{K}{}^{L}b_{L},\ \ \ \Lambda _{K}{}^{L}\in GL(4,%
\mathbb{R})\ 
\end{equation*}%
the covariant derivative of the vector 
\begin{equation*}
D_{\mu }V^{K}=\partial _{\mu }V^{K}+\Gamma ^{K}{}_{L\mu }V^{L}.\ \ \ 
\end{equation*}%
becomes%
\begin{equation*}
D_{\mu }V^{\prime K}=\partial _{\mu }V^{K}+\Gamma ^{\prime K}{}_{L\mu
}V^{L}.\ \ \ 
\end{equation*}%
On account of the transformation rules 
\begin{equation*}
\Gamma ^{\prime K}{}_{I\mu }=\left( \Lambda ^{-1}\right) _{L}{}^{K}\Gamma
^{L}{}_{J\mu }\Lambda _{I}{}^{J}+\left( \Lambda ^{-1}\right)
_{J}{}^{K}\partial _{\mu }\Lambda _{I}{}^{J},
\end{equation*}%
the covariant derivative $D_{\mu }V^{\prime K}$ can be transform to%
\begin{equation*}
D_{\mu }V^{\prime K}=0,
\end{equation*}%
by a appropriate transformation $\Lambda _{K}{}^{L}$.,

By the same way, the covariant derivatives of any $(k,l)$ type tensor $%
T^{I_{1}\cdots I_{k}}{}_{J_{1}\cdots J_{l}}$, $D_{\mu }T^{I_{1}\cdots
I_{k}}{}_{J_{1}\cdots J_{l}}$ can vanishes under some base transformation
and, therefore, form a special kind of tensor which could be called \textsl{%
composite tensors}. A composite tensor is composed of two parts, a partial
derivative and terms of the connection, none of these terms alone are
tensors.

\subsection{Torsion and curvature}

Suppose that $\left\{ b_{I}\right\} $ spans a Lie algeba, the base vectors $%
b_{I}$ satisfy the commutation rule%
\begin{equation}
\lbrack b_{K},b_{L}]=f^{I}{}_{KL}b_{I},
\end{equation}%
where%
\begin{equation}
f^{I}{}_{KL}=b_{K}{}^{\mu }b_{L}{}^{\nu }\left( \partial _{\nu }b^{I}{}_{\mu
}-\partial _{\mu }b^{I}{}_{\nu }\right) .
\end{equation}%
If $f^{I}{}_{KL}=0,$the basis $\left\{ b_{K}\right\} $ is holonomic and if $%
f^{I}{}_{KL}\neq 0$, $\left\{ b_{K}\right\} $ is\ anholonomic. The
coordinate basis $\left\{ \partial _{\mu }\right\} $ is holonomic,
obviously. The dual expression of the commutation relation above is the
Cartan structure equation

\begin{equation}
db^{I}=\frac{1}{2}f^{I}{}_{JK}b^{J}\wedge b^{K}=\frac{1}{2}(\partial _{\nu
}b^{I}{}_{\mu }-\partial _{\mu }b^{I}{}_{\nu })dx^{%
\mu
}\wedge dx^{\nu }.
\end{equation}

\textsl{Definition: } The two-form%
\begin{equation}
db^{I}=\frac{1}{2}T^{I}{}_{\mu \nu }dx^{%
\mu
}\wedge dx^{\nu }\text{ \ \ \ }
\end{equation}%
with%
\begin{equation}
T^{I}{}_{\mu \nu }=\partial _{\nu }b^{I}{}_{\mu }-\partial _{\mu
}b^{I}{}_{\nu },
\end{equation}%
is called the \textsl{Cartan} \textsl{torsion of the basis} $\left\{
b_{K}\right\} $. It is simply determined by the structure constants, $%
f^{I}{}_{KL}$, of the Lie algeba.

Now we have 
\begin{equation}
f^{I}{}_{KL}=b_{K}{}^{\mu }b_{L}{}^{\nu }T^{I}{}_{\mu \nu },
\end{equation}%
and%
\begin{equation}
\lbrack b_{K},b_{L}]=b_{K}{}^{\mu }b_{L}{}^{\nu }T^{I}{}_{\mu \nu }b_{I}.
\end{equation}%
The Cartan torsion is identified with the anholonomy of the frame.
Therefore, the holonomic basis $\left\{ \partial _{\mu }\right\} $\ is
torsion free.

Since the basis $\{b_{K}\}$ are fields i.e. $b_{K}=b_{K}\left( x\right) $,
the ordinary derivative $\partial _{\mu }$ has to be substituted by the
covariant derivative $D_{\mu }$, the Cartan torsion $\left( 18\right) $ of $%
\left\{ b_{I}\right\} $ now becomes 
\begin{equation}
T^{I}{}_{\mu \nu }=D_{\nu }b^{I}{}_{\mu }-D_{\mu }b^{I}{}_{\nu }=\partial
_{\nu }b^{I}{}_{\mu }-\partial _{\mu }b^{I}{}_{\nu }+\Gamma ^{I}{}_{J\nu
}b^{J}{}_{\mu }-\Gamma ^{I}{}_{J\mu }b^{J}{}_{\nu },
\end{equation}%
which is called the\textsl{\ torsion of the basis} $\{b_{K}\}$.

Recalling $\left( 11\right) $ we have%
\begin{equation}
T^{\rho }{}_{\mu \nu }=b_{I}{}^{\rho }T^{I}{}_{\mu \nu }=\Gamma ^{\rho
}{}_{\nu \mu }-\Gamma ^{\rho }{}_{\mu \nu }.
\end{equation}%
Usually, $T^{\rho }{}_{\mu \nu }$ is called the\textsl{\ torsion of the of
the connection }$\Gamma ^{\rho }{}_{\mu \nu }$\textsl{, }although\textsl{\
it essentially belongs to the frame} $\{b_{K}\}$. The torsion essentially is
the external derivative of the frame \{$b^{I}$\}. It reflect a character of $%
\{b_{K}\}$, i.e. the anholonomy.

Using $\left( 4\right) $ and $\left( 8\right) $ we can obtain%
\begin{equation}
\nabla _{\mu }\nabla _{\nu }V^{\alpha }-\nabla _{\nu }\nabla _{\mu
}V^{\alpha }=R^{\alpha }{}_{\lambda \mu \nu }V^{\lambda }+T^{\lambda
}{}_{\mu \nu }\nabla _{\lambda }V^{\alpha }\ \ \ \ 
\end{equation}%
and%
\begin{equation}
D_{\mu }D_{\nu }V^{I}-D_{\nu }D_{\mu }V^{I}=R^{I}{}_{J\mu \nu }V^{J},\ \ \ \ 
\end{equation}%
where%
\begin{equation}
R^{I}{}_{J\mu \nu }=R_{\Gamma }^{I}{}_{J\mu \nu }=\partial _{\mu }\Gamma
^{I}{}_{J\nu }-\partial _{\nu }\Gamma ^{I}{}_{J\mu }+\Gamma ^{I}{}_{K\mu
}\Gamma ^{K}{}_{J\nu }-\Gamma ^{I}{}_{K\nu }\Gamma ^{K}{}_{J\mu },\ \ 
\end{equation}%
and%
\begin{equation}
R^{\alpha }{}_{\lambda \mu \nu }=R_{\Gamma }^{\alpha }{}_{\beta \mu \nu
}=\partial _{\mu }\Gamma ^{\alpha }{}_{\beta \nu }-\partial _{\nu }\Gamma
^{\alpha }{}_{\beta \mu }+\Gamma ^{\alpha }{}_{\rho \mu }\Gamma ^{\rho
}{}_{\beta \nu }-\Gamma ^{\alpha }{}_{\rho \nu }\Gamma ^{\rho }{}_{\beta \mu
},
\end{equation}%
are called the\ curvature of the connections $\Gamma ^{I}{}_{K\mu }$ and $%
\Gamma ^{\alpha }{}_{\beta \mu }$, respectively, which are related each
other by 
\begin{equation}
R^{\alpha }{}_{\beta \mu \nu }=b_{I}{}^{\alpha }b^{J}{}_{\beta
}R^{I}{}_{J\mu \nu }.
\end{equation}

In contrary to the holonimic basis $\left\{ \partial _{\mu }\right\} $, the
basis $\left\{ \nabla _{\mu }\right\} $ and $\left\{ D_{\mu }\right\} $ are
anholonomic as indicated by $\left( 23\right) $ and $\left( 24\right) $. $%
T^{\lambda }{}_{\mu \nu }$, $R^{\alpha }{}_{\lambda \mu \nu }$ and $%
R^{I}{}_{J\mu \nu }$ are the the measure of the anholonomy\textsl{.}

\subsubsection{Remarks}

1. We should carefully recognize different characters under different
transformations of a same geometry object. For example, $\Gamma ^{I}{}_{J\mu
}$ is an 1-form under a coordinate transformation, but is a connection under
a frame transformation not a tensor. $\Gamma ^{\alpha }{}_{\beta \mu }$ is
an 1-form and a connection not a tensor under a coordinate transformation
but is a scalar under a frame transformation. $T^{\lambda }{}_{\mu \nu }$ is
an 1 rank contravariant and 2 rank covariant skew-symmetric tensor under a
coordinate transformation, and a scalar under a frame transformation. $%
R^{I}{}_{J\mu \nu }$ is a 2-form under a coordinate transformation and a 2
rank skew-symmetric tensor under a frame transformation. $R^{\lambda
}{}_{\tau \mu \nu }$ is an 1 rank contravariant and 3 rank covariant tensor
under a coordinate transformation, and a scalar under a frame transformation.

2. All the transformations $b_{I}\rightarrow b_{I}^{\prime }$\ take place on
the tangent space $T_{{\Large x}}M$. Especially, although, the coordinate
transformation $x^{\mu }\rightarrow x^{\prime \mu }$ take place on the
manifold $M$, the associated linear transformations $\partial _{\mu
}\rightarrow \partial _{\mu }^{\prime }$\ take place on the tangent space $%
T_{{\Large x}}M$. Therefore, the coordinate bases $\left\{ \partial _{\mu
}\right\} $ are the special cases of the frames $\left\{ b_{I}\right\} $,
the connections $\Gamma ^{\rho }{}_{\nu \mu }$ are also the special cases of
the connections $\Gamma ^{I}{}_{J\mu }$. The covariant derivatives $\nabla
_{\mu }$ are the special cases of the covariant derivatives $D_{\mu }$. $%
\nabla _{\mu }$ acts on coordinate indices $\mu ,\nu ,...$\ only and is
called the "external" derivative usually, although $\left\{ \partial _{\mu
}\right\} $ lie in $T_{{\Large x}}M$, while $D_{\mu }$ acts on indices $%
I,J,...$ and is called the "internal" derivative usually. In this sense, 
\textsl{the distinction of the "external" and the "internal" is not exact
well}. The torsion 
\begin{equation*}
T^{\rho }{}_{\mu \nu }=\Gamma ^{\rho }{}_{\nu \mu }-\Gamma ^{\rho }{}_{\mu
\nu }
\end{equation*}%
is the special cases of the torsion 
\begin{equation*}
T^{I}{}_{\mu \nu }=D_{\nu }b^{I}{}_{\mu }-D_{\mu }b^{I}{}_{\nu }.
\end{equation*}%
In some sense, $R^{I}{}_{J\mu \nu }$ can be called the "torsion" of the
connection $\Gamma ^{I}{}_{J\mu }$, while $R^{\alpha }{}_{\beta \mu \nu }$
can be called the "torsion" of the connection $\Gamma ^{\alpha }{}_{\beta
\mu }$. The torsion $T^{\rho }{}_{\mu \nu }$ belongs to the frame $\{b_{K}\}$
rather than the \textsl{connection }$\Gamma ^{\rho }{}_{\mu \nu }$. \textsl{%
The phrase "torsion" means anholonomy or noncommutativity}.

3. \textsl{It should be emphasized that neither }$T^{I}{}_{\mu \nu }$\textsl{%
\ nor }$R^{I}{}_{J\mu \nu }$\textsl{\ belongs to the manifold. The torsion }$%
T^{I}{}_{\mu \nu }$\textsl{\ belongs to the basis }$\left\{ b_{I}\right\} $%
\textsl{, while the curvature }$R^{I}{}_{J\mu \nu }$\textsl{\ belongs to the
connection }$\left\{ \Gamma ^{I}{}_{J\mu }\right\} $. Therefore, we can only
say "a curved connection" or "a flat connection" not "a curved manifold" or
"a flat manifold". We can only say "the torsion of a frame" not "the torsion
of a manifold". There is no such things as curvature or torsion of spacetime
[30, {\small p302}] if the spacetime is considered as a manifold.

4. The frame vector (or 1-form) fields on a manifold can be classified as
tow types: holonomic and anholonomic. The anholonomic vector fields can be
classified into two categories further: the first category can form a Lie
algebra, while the second category can not. For example, when $R^{I}{}_{J\mu
\nu }\neq 0$, either \{$D_{\mu }$\} or \{$\nabla _{\mu }$\} do not form a
Lie algebra.

5.{\LARGE \ }\textbf{Flat connection}

Curvature is a \textsl{composite tensors, }it can vanish under appropriate
transformation\textsl{.} When 
\begin{equation*}
R^{I}{}_{J\mu \nu }=0
\end{equation*}%
the connection $\Gamma ^{L}{}_{J\mu }$ is called a\textsl{\ flat connection
or an uniform (homogeneous) connection}. When 
\begin{equation*}
R^{I}{}_{J\mu \nu }\neq 0
\end{equation*}%
the connection $\Gamma ^{L}{}_{J\mu }$ is called a\textsl{\ curved
connection or a ununiform (unhomogeneous) connection}.

If%
\begin{equation*}
\Gamma ^{L}{}_{J\mu }=0,
\end{equation*}%
(5) leads to%
\begin{equation}
\Gamma ^{\prime K}{}_{I\mu }=\left( \Lambda ^{-1}\right) _{J}{}^{K}\partial
_{\mu }\Lambda _{I}{}^{J}.\ \ \ 
\end{equation}%
A simple calculation shows that the curvature of $\Gamma ^{\prime K}{}_{I\mu
}$ vanishes, so $\Gamma ^{\prime K}{}_{I\mu }$ is flat connection. This
means that a flat connection appears with a local $GL(4,\mathbb{R})$
transformation $\Lambda _{I}{}^{J}\left( x\right) $.

As specific examples the flat connection [5]%
\begin{equation}
\Gamma ^{\prime \lambda }{}_{\nu \mu }=\frac{\partial x^{\prime \lambda }}{%
\partial x^{\tau }}\frac{\partial ^{2}x^{\tau }}{\partial x^{\prime \nu
}\partial x^{\prime \mu }}\ .
\end{equation}%
appears with a coordinate transformation $x^{\mu }\rightarrow x^{\prime \mu
} $, the flat connections

\begin{equation}
\Gamma ^{\rho }{}_{\mu \nu }=b_{I}{}^{\rho }\partial _{\mu }b^{I}{}_{\nu }\
.\ \ \ 
\end{equation}%
and%
\begin{equation}
\Gamma ^{I}{}_{J\mu }=b^{I}{}_{\nu }\partial _{\mu }b_{J}{}^{\nu }\ .\ \ \ 
\end{equation}%
\ appear with transformations from the general basis $\left\{ b_{I}\right\} $
to the coordinate basis $\left\{ \partial _{\mu }\right\} $, and from the
coordinate $\left\{ \partial _{\mu }\right\} $ to the general basis $\left\{
b_{I}\right\} $, respectively.

As a complex tensor, the curvature distinguishes from other tensors which
can not vanish under any transformation\textsl{.} \textsl{The curvature
belongs to a connection rather than the manifold. There can exists \textbf{%
not only} one} \textsl{connection and curvature on a manifold. }

\subsection{The translation group and the translation connection}

In the tangent space $T_{x}M$, a vector $V$ corresponds to a point $y$, the
components $V^{I}$ of the vector correspond to the coordinates $y^{I}$ of
the point on $T_{x}M$.\ In addition to the basis transformations there is
another kind of transformations, the translation on $T_{x}M$ [33],%
\begin{equation}
y^{I}\rightarrow y^{\prime I}=y^{I}+\varepsilon ^{I}\left( x^{\mu }\right) .
\end{equation}

The derivatives $\partial _{\mu }=\partial /\partial x^{\mu }$ and $\partial
_{I}=\partial /\partial y^{I}$ are related each other by

\begin{equation*}
\partial _{\mu }=(\partial _{\mu }y^{I})\partial _{I},\partial
_{I}=(\partial _{I}x^{\mu })\partial _{\mu }.
\end{equation*}%
Under the translations $y^{I}\rightarrow y^{\prime I}$, a vector $V^{I}$
transform as%
\begin{equation*}
V^{I}\rightarrow V^{\prime I}=V^{I},
\end{equation*}%
while the derivative $\partial _{\mu }V^{I}$, transforms according to the
rule%
\begin{equation*}
\partial _{\mu }V^{\prime I}=\partial _{\mu }V^{I}+\partial _{\mu
}\varepsilon ^{J}\partial _{J}V^{I}.
\end{equation*}%
If there is a \textsl{translation connection} $B^{I{}}{}_{\mu }$ which
defines the \textsl{covariant derivative}%
\begin{equation}
h_{\mu }^{\left( 0\right) }V^{I}=\left( \partial _{\mu }y^{J}+B^{J{}}{}_{\mu
}\right) \partial _{J}V^{I},
\end{equation}%
and satisfies the transformation rule, 
\begin{equation}
B^{\prime J{}}{}_{\mu }=B^{J{}}{}_{\mu }-\partial _{\mu }\varepsilon ^{J},
\end{equation}%
then\ we have%
\begin{equation*}
\partial _{\mu }y^{\prime J}\partial _{J}V^{\prime I}+B^{\prime J{}}{}_{\mu
}\partial _{J}V^{\prime I}=\partial _{\mu }y^{J}\partial
_{J}V^{I}+B^{J{}}{}_{\mu }\partial _{J}V^{I},
\end{equation*}%
i.e.%
\begin{equation}
h_{\mu }^{\prime \left( 0\right) }V^{\prime I}=h_{\mu }^{\left( 0\right)
}V^{I},
\end{equation}%
which means that $h_{\mu }^{\left( 0\right) }V^{I}$ is covariant with $V^{I}$
under the translation.

\textsl{Considering the two transformations simultaneously}, the linear
transformation and the translation, 
\begin{equation}
y^{I}\rightarrow y^{\prime I}=\Lambda ^{I}{}_{J}y^{J}+\varepsilon ^{I},\text{
\ \ \ }
\end{equation}%
we must take $D_{\mu }y^{I}$ instead of $\partial _{\mu }y^{I}$, and then
have%
\begin{eqnarray}
h_{\mu }V^{I} &=&\left( D_{\mu }y^{J}+B^{J{}}{}_{\mu }\right) \partial
_{J}V^{I}=\left( \partial _{\mu }y^{J}+\Gamma ^{J}{}_{K\mu
}y^{K}+B^{J{}}{}_{\mu }\right) \partial _{J}V^{I}  \notag \\
&=&\left( h^{\left( 0\right) J{}}{}_{\mu }+\Gamma ^{J}{}_{K\mu }y^{K}\right)
\partial _{J}V^{I}  \notag \\
&=&h^{J{}}{}_{\mu }\partial _{J}V^{I},\text{ \ \ \ \ \ \ \ \ \ \ \ \ \ \ \ \
\ \ \ \ \ \ \ \ \ \ \ \ \ \ \ }
\end{eqnarray}%
where%
\begin{equation}
h^{I{}}{}_{\mu }:=h^{\left( 0\right) I{}}{}_{\mu }+\Gamma ^{I}{}_{J\mu
}y^{J}=\partial _{\mu }y^{I}+\Gamma ^{I}{}_{J\mu }y^{J}+B^{I{}}{}_{\mu }.%
\text{ \ \ }
\end{equation}%
A local translation transforms trivial\ frame $\partial _{\mu }y^{I}$\ to a
nontrivial frame $h^{\left( 0\right) I}{}{}_{\mu }$\ and introduces a gauge
potential (translation connection) $B^{a}{}_{\mu }$. At the same time, a
local frame transformation introduces a gauge potential (affine connection) $%
\Gamma ^{I}{}_{J\mu }$. As thus, \textsl{the general frame }$h^{I{}}{}_{\mu
} $\textsl{\ is given by a combination of the connections }$\Gamma
^{I}{}_{J\mu }$\textsl{\ and }$B^{I{}}{}_{\mu }$\textsl{.}

The covariant derivative operators\textsl{\ }%
\begin{equation}
h_{\mu }=h^{I{}}{}_{\mu }\partial _{I}\text{ \ \ \ }
\end{equation}%
constitute an anholonomic basis with the dual 
\begin{equation}
h^{\mu }=h_{I}{}^{\mu }dy^{I},
\end{equation}%
satisfying%
\begin{equation}
h^{\mu }\left( h_{\nu }\right) =h_{I}{}^{\mu }h^{J{}}{}_{\nu }dx^{I}\left(
\partial _{J}\right) =h_{I}{}^{\mu }h^{J{}}{}_{\nu }\delta
_{J}^{I}=h_{I}{}^{\mu }h^{I{}}{}_{\nu }=\delta _{\nu }^{\mu }.
\end{equation}%
The Eq. $h_{\mu }=h^{I{}}{}_{\mu }\partial _{I}$ leads to%
\begin{equation}
\partial _{I}=h_{I}{}^{\mu }h_{\mu },
\end{equation}%
and 
\begin{equation}
\left[ h_{\mu },h_{\nu }\right] =f^{\lambda {}}{}_{\mu }{}_{\nu }h_{\lambda }
\end{equation}%
with%
\begin{equation}
f^{\lambda {}}{}_{\mu }{}_{\nu }=h^{I{}}{}_{\mu }h^{J}{}_{\nu }\left(
\partial _{J}h_{I}{}^{\lambda }-\partial _{I}h_{J}{}^{\lambda }\right) ,
\end{equation}%
where $h_{I}{}^{\mu }$ is the reciprocal of $h^{I{}}{}_{\mu }$.

The two-form%
\begin{equation}
dh^{\lambda }=\frac{1}{2}T^{\lambda }{}_{\mu \nu }dx^{%
\mu
}\wedge dx^{\nu }\text{ \ \ \ }
\end{equation}%
with%
\begin{equation}
T^{\lambda }{}_{\mu \nu }=h_{I}{}^{\lambda }T^{I}{}_{\mu \nu }\text{ \ \ \ }
\end{equation}%
is the \textsl{torsion of the basis} $\left\{ h_{\mu }\right\} $, where%
\begin{equation}
T^{I}{}_{\mu \nu }=D_{\nu }h^{I}{}_{\mu }-D_{\mu }h^{I}{}_{\nu }=\partial
_{\nu }h^{I}{}_{\mu }-\partial _{\mu }h^{I}{}_{\nu }+\Gamma ^{I}{}_{J\nu
}h^{J}{}_{\mu }-\Gamma ^{I}{}_{J\mu }h^{J}{}_{\nu }.
\end{equation}

Using%
\begin{equation*}
h^{I{}}{}_{\mu }:=\partial _{\mu }y^{I}+\Gamma ^{I}{}_{J\mu
}y^{J}+B^{I{}}{}_{\mu }.\text{ \ \ }
\end{equation*}%
we have%
\begin{equation}
T^{I}{}_{\mu \nu }=-R^{I}{}_{J\mu \nu }y^{J}+D_{\nu }B^{I{}}{}_{\mu }-D_{\mu
}B^{I{}}{}_{\nu }.\ \ \ \ 
\end{equation}%
This formula relates the curvature $R^{I}{}_{J\mu \nu }$ to the torsion $%
T^{I}{}_{\mu \nu }$.

It should be emphasized that the torsion $T^{I}{}_{\mu \nu }$\ belongs to
the basis $\left\{ h_{\mu }\right\} $, while the curvature $R^{I}{}_{J\mu
\nu }$\ belongs to the connection $\left\{ \Gamma ^{I}{}_{J\mu }\right\} $.
The torsion $T^{I}{}_{\mu \nu }$\ belongs to the translation transformation $%
T^{4}$, while the curvature $R^{I}{}_{J\mu \nu }$\ belongs to the linear
transformation $GL\left( 4,\mathbb{R}\right) $

Since \textsl{the general frame }$h^{I{}}{}_{\mu }$\textsl{\ is a
combination of the connections }$\Gamma ^{I}{}_{J\mu }$\textsl{\ and }$%
B^{I{}}{}_{\mu }$\textsl{, it's torsion }$T^{I}{}_{\mu \nu }$\textsl{\
includes naturally both the "torsion\textquotedblleft\ }$R^{I}{}_{J\mu \nu }$%
\textsl{\ of }$\Gamma ^{I}{}_{J\mu }$\textsl{\ and the "torsion" }$%
F^{I}{}_{\mu \nu }=D_{\nu }B^{I{}}{}_{\mu }-D_{\mu }B^{I{}}{}_{\nu }$\textsl{%
\ of }$B^{I{}}{}_{\mu }$\textsl{. }There are two special cases:

i) If $\Gamma ^{I}{}_{J\nu }=0$, we have 
\begin{equation}
T^{I}{}_{\mu \nu }=\partial _{\nu }B^{I{}}{}_{\mu }-\partial _{\mu
}B^{I{}}{}_{\nu }=-F^{I}{}_{\mu \nu }.\ \ \ \ 
\end{equation}%
The anholonomy of $h^{I{}}{}_{\mu }:=\partial _{\mu }y^{I}+B^{I{}}{}_{\mu }$
is due to $B^{I{}}{}_{\mu }$.\ 

ii) If $B^{I{}}{}_{\mu }=0$, we have%
\begin{equation}
T^{I}{}_{\mu \nu }=-R^{I}{}_{J\mu \nu }y^{J}.
\end{equation}%
The torsion $T^{I}{}_{\mu \nu }$ of $h^{I{}}{}_{\mu }$ is proportional to
the curvature $R^{I}{}_{J\mu \nu }$\textrm{\textbf{. }}The anholonomy of $%
h^{I{}}{}_{\mu }:=\partial _{\mu }y^{I}+\Gamma ^{I}{}_{J\mu }y^{J}$ is due
to $\Gamma ^{I}{}_{J\mu }$.

\subsection{Metric geometry}

\ In addition to the affine structure, the manifold $M$ can be endow with a
metric structure by a tensor field of type ($0,2$), i.e. the metric tensor $%
\mathbf{g}$.{\LARGE \ }The tangent space $T_{x}M$ of the manifold is a
vector space. The metric tensor $\mathbf{g}$ defines the \textbf{scalar
product} $\mathbf{g}\left( u,v\right) $ of two vectors $u$ and $v$ and the
modulus $\mathbf{g}\left( v,v\right) $ of a vector $v$. For the bases $%
\left\{ b_{I}\right\} $, $\left\{ \partial _{\mu }\right\} $, respectively,
we have 
\begin{equation}
\mathbf{g}\left( b_{I},b_{J}\right) =g_{IJ}\left( x\right) ,
\end{equation}%
and%
\begin{equation}
\mathbf{g}\left( \partial _{\mu },\partial _{\nu }\right) =g_{\mu \nu
}\left( x\right) .
\end{equation}%
If $g_{IJ}=$ diag $\left( -1,1,1,1\right) =\eta _{IJ}$, \{$b_{I}=\partial
_{I}$\} is called a Lorentz frame or a \textsl{tetrad}.{\LARGE \ }Therefore
the metric tensor can pick out tetrads from general frames on $T_{x}M$.

In the case 
\begin{equation*}
h^{I{}}{}_{\mu }=\partial _{\mu }y^{I}+\Gamma ^{I}{}_{J\mu
}y^{J}+B^{I{}}{}_{\mu },
\end{equation*}%
it gives%
\begin{equation}
\mathbf{g}\left( h_{\mu },h_{\nu }\right) =\mathbf{g}\left( h^{I{}}{}_{\mu
}\partial _{I},h^{J}{}_{\nu }\partial _{J}\right) =h^{I{}}{}_{\mu
}h^{J}{}_{\nu }g_{IJ}\left( x\right) =g_{\mu \nu }^{\left( h\right) }\left(
x\right) ,
\end{equation}%
with%
\begin{eqnarray}
g_{\mu \nu }^{\left( h\right) }\left( x\right) &=&g_{IJ}(\partial _{\mu
}y^{I}\partial _{\nu }y^{J}+\Gamma ^{I}{}_{K\mu }\Gamma ^{J}{}_{L\nu
}y^{K}y^{L}+\partial _{\nu }y^{J}\Gamma ^{I}{}_{K\mu }y^{K}  \notag \\
&&+\partial _{\mu }y^{I}\Gamma ^{J}{}_{L\nu }y^{L}+\partial _{\nu
}y^{J}B^{I{}}{}_{\mu }+\partial _{\mu }y^{I}B^{J{}}{}_{\nu }  \notag \\
&&+\Gamma ^{J}{}_{L\nu }y^{L}B^{I{}}{}_{\mu }+\Gamma ^{I}{}_{K\mu
}y^{K}B^{J{}}{}_{\nu }+B^{I{}}{}_{\mu }B^{J{}}{}_{\nu })\text{ \ }
\end{eqnarray}%
Like the frame $h^{I{}}{}_{\mu }$, $g_{\mu \nu }^{\left( h\right) }$ is also
the combination of the connections $B^{I{}}{}_{\mu }$ and $\Gamma
^{I}{}_{J\mu }$. However,

i) when $B^{I{}}{}_{\mu }=0$, it depends on only $\Gamma ^{I}{}_{J\mu }$:%
\begin{equation}
g_{\mu \nu }^{\left( h\right) }\left( x\right) =\left( \partial _{\mu
}y^{I}\partial _{\nu }y^{J}+\Gamma ^{I}{}_{K\mu }\Gamma ^{J}{}_{L\nu
}y^{K}y^{L}+\partial _{\nu }y^{J}\Gamma ^{I}{}_{K\mu }y^{K}+\partial _{\mu
}y^{I}\Gamma ^{J}{}_{L\nu }y^{L}\right) g_{IJ},
\end{equation}

ii) when $\Gamma ^{I}{}_{J\mu }=0$, it depends on only $B^{I{}}{}_{\mu }$:%
\begin{equation}
g_{\mu \nu }^{\left( h\right) }\left( x\right) =\left( \partial _{\mu
}y^{I}\partial _{\nu }y^{J}+\partial _{\nu }y^{J}B^{I{}}{}_{\mu }+\partial
_{\mu }y^{I}B^{J{}}{}_{\nu }+B^{I{}}{}_{\mu }B^{J{}}{}_{\nu }\right) g_{IJ},
\end{equation}

iii) when $B^{I{}}{}_{\mu }=0$, $\Gamma ^{I}{}_{J\mu }=0$,%
\begin{equation}
g_{\mu \nu }^{\left( h\right) }=g_{IJ}\partial _{\mu }y^{I}\partial _{\nu
}y^{J}\text{ \ \ }.
\end{equation}

If $g_{IJ}=$ diag $\left( -1,1,1,1\right) =\eta _{IJ}$, \{$\partial
_{I}=b_{I}$\} is a Lorentz frame or a \textsl{tetrad}, the transformations \{%
$\Lambda _{K}{}^{L}$\} constitute a Lorentz group $SO(1,3)$. Now we get a
sub-bundle i.e. the bundle of orthonormal frames. It consists of orthonormal
bases $\left\{ \partial _{I}\right\} $ with respect to $\eta _{IJ}$, with
the Lorentz group $SO(1,3)$ as the structure group. In this case,%
\begin{equation*}
h^{I{}}{}_{\mu }=\partial _{\mu }y^{I}+\Gamma ^{I}{}_{J\mu
}y^{J}+B^{I{}}{}_{\mu }
\end{equation*}%
will be denoted as%
\begin{equation}
e^{I{}}{}_{\mu }=\partial _{\mu }y^{I}+\omega ^{I}{}_{J\mu
}y^{J}+B^{I{}}{}_{\mu }\text{ \ \ }
\end{equation}%
with the Lorentz connection $\omega ^{I}{}_{J\mu }$. The corresponding
curvature 
\begin{equation}
R_{\omega }^{I}{}_{J\mu \nu }=\partial _{\mu }\omega ^{I}{}_{J\nu }-\partial
_{\nu }\omega ^{I}{}_{J\mu }+\omega ^{I}{}_{K\mu }\omega ^{K}{}_{J\nu
}-\omega ^{I}{}_{K\nu }\omega ^{K}{}_{J\mu }\text{ \ \ \ }
\end{equation}%
is called the \textsl{Lorentz curvature} which is related to the affine
curvature $R_{\Gamma }^{\alpha }{}_{\beta \mu \nu }$ by 
\begin{equation}
R_{\Gamma }^{\alpha }{}_{\beta \mu \nu }=e_{I}{}^{\alpha }e^{J}{}_{\beta
}R_{\omega }^{I}{}_{J\mu \nu },
\end{equation}%
with 
\begin{equation}
\Gamma ^{\alpha }{}_{\mu \nu }=e_{I}{}^{\alpha }\partial _{\mu }e^{I}{}_{\nu
}+e_{I}{}^{\alpha }\omega ^{I}{}_{J\mu }e^{J}{}_{\nu }.
\end{equation}%
\bigskip

The covariant derivatives $\nabla _{\mu }g_{AB}$ of the metric $g_{AB}$
define the non-metricity%
\begin{equation}
Q_{\mu AB}\equiv D_{\mu }g_{AB}=\partial _{\mu }g_{AB}-\Gamma ^{C}{}_{A\mu
}g_{CB}-\Gamma ^{C}{}_{B\mu }g_{AC}.
\end{equation}%
For the coordinate basis $\left\{ \partial _{\mu }\right\} $, non-metricity
is 
\begin{equation}
Q_{\mu \alpha \beta }\equiv \nabla _{\mu }g_{\alpha \beta }=\partial _{\mu
}g_{\alpha \beta }-\Gamma ^{\rho }{}_{\alpha \mu }g_{\rho \beta }-\Gamma
^{\rho }{}_{\beta \mu }g_{\alpha \rho }.
\end{equation}

Once a metric field is defined on the manifold, the affine connection $%
\Gamma {}_{\mu \alpha \beta }=g_{\mu \nu }\Gamma ^{\nu {}}{}_{\alpha \beta }$
can be decomposed into three parts:%
\begin{equation}
\Gamma {}_{\mu \alpha \beta }=\left\{ _{\alpha \mu \beta }\right\} +K_{\mu
\alpha \beta }+L_{\mu \alpha \beta },
\end{equation}%
where%
\begin{eqnarray}
\left\{ _{\alpha \mu \beta }\right\} &=&\frac{1}{2}\left( \partial _{\alpha
}g_{\mu \beta }+\partial _{\beta }g_{\alpha \mu }-\partial _{\mu }g_{\alpha
\beta }\right) , \\
K_{\mu \alpha \beta } &=&\frac{1}{2}\left( T_{\beta \alpha \mu }+T_{\alpha
\beta \mu }-T_{\mu \alpha \beta }\right) ,\text{ \ \ \ \ \ \ } \\
L_{\mu \alpha \beta } &=&\frac{1}{2}\left( Q_{\mu \alpha \beta }-Q_{\alpha
\mu \beta }-Q_{\beta \alpha \mu }\right) ,\text{ \ \ }
\end{eqnarray}%
are referred as the Levi-Civita connection, the contortion and the
disformation, separately.

The equations%
\begin{equation*}
R_{\Gamma }^{\alpha }{}_{\beta \mu \nu }=\partial _{\mu }\Gamma ^{\alpha
}{}_{\beta \nu }-\partial _{\nu }\Gamma ^{\alpha }{}_{\beta \mu }+\Gamma
^{\alpha }{}_{\rho \mu }\Gamma ^{\rho }{}_{\beta \nu }-\Gamma ^{\alpha
}{}_{\rho \nu }\Gamma ^{\rho }{}_{\beta \mu },
\end{equation*}%
and

\begin{equation*}
\Gamma ^{\mu }{}_{\alpha \beta }=\left\{ _{\alpha }{}^{\mu }{}_{\beta
}\right\} +K^{\mu }{}_{\alpha \beta }+L^{\mu }{}_{\alpha \beta }
\end{equation*}%
leads to%
\begin{equation}
R_{\Gamma }^{\alpha }{}_{\beta \mu \nu }=R_{\left\{ {}\right\} }^{\alpha
}{}_{\beta \mu \nu }+R_{K}^{\alpha }{}_{\beta \mu \nu }+R_{L}^{\alpha
}{}_{\beta \mu \nu }+\Theta ^{\alpha }{}_{\beta \mu \nu }\text{ \ \ }
\end{equation}%
where%
\begin{eqnarray}
R_{\left\{ {}\right\} }^{\alpha }{}_{\beta \mu \nu } &=&\partial _{\mu
}\left\{ _{\beta }{}^{\alpha }{}_{\nu }\right\} -\partial _{\nu }\left\{
_{\beta }{}^{\alpha }{}_{\mu }\right\} +\left\{ _{\rho }{}^{\alpha }{}_{\mu
}\right\} \left\{ _{\beta }{}^{\rho }{}_{\nu }\right\} -\left\{ _{\rho
}{}^{\alpha }{}_{\nu }\right\} \left\{ _{\beta }{}^{\rho }{}_{\mu }\right\} ,
\\
R_{K}^{\alpha }{}_{\beta \mu \nu } &=&\partial _{\mu }K^{\alpha }{}_{\beta
\nu }-\partial _{\nu }K^{\alpha }{}_{\beta \mu }+K^{\alpha }{}_{\rho \mu
}K^{\rho }{}_{\beta \nu }-K^{\alpha }{}_{\rho \nu }K^{\rho }{}_{\beta \mu },%
\text{ \ \ \ \ \ } \\
R_{L}^{\alpha }{}_{\beta \mu \nu } &=&\partial _{\mu }L^{\alpha }{}_{\beta
\nu }-\partial _{\nu }L^{\alpha }{}_{\beta \mu }+L^{\alpha }{}_{\beta \mu
}L^{\rho }{}_{\beta \nu }-L^{\alpha }{}_{\rho \nu }L^{\rho }{}_{\beta \mu },%
\text{ \ \ \ \ \ \ \ }
\end{eqnarray}%
\begin{eqnarray}
\Theta ^{\alpha }{}_{\beta \mu \nu } &=&\left\{ _{\rho }{}^{\alpha }{}_{\mu
}\right\} K^{\rho }{}_{\beta \nu }-\left\{ _{\rho }{}^{\alpha }{}_{\nu
}\right\} K^{\rho }{}_{\beta \mu }+K^{\alpha }{}_{\rho \mu }\left\{ _{\beta
}{}^{\rho }{}_{\nu }\right\} -K^{\alpha }{}_{\rho \nu }\left\{ _{\beta
}{}^{\rho }{}_{\mu }\right\}  \notag \\
&&+\left\{ _{\rho }{}^{\alpha }{}_{\mu }\right\} L^{\rho }{}_{\beta \nu
}-\left\{ _{\rho }{}^{\alpha }{}_{\nu }\right\} L^{\rho }{}_{\beta \mu
}+L^{\alpha }{}_{\beta \mu }\left\{ _{\beta }{}^{\rho }{}_{\nu }\right\}
-L^{\alpha }{}_{\rho \nu }\left\{ _{\beta }{}^{\rho }{}_{\mu }\right\} 
\notag \\
&&+K^{\alpha }{}_{\rho \mu }L^{\rho }{}_{\beta \nu }-K^{\alpha }{}_{\rho \nu
}L^{\rho }{}_{\beta \mu }+L^{\alpha }{}_{\beta \mu }K^{\rho }{}_{\beta \nu
}-L^{\alpha }{}^{\rho }{}_{\beta \mu \rho \nu }K.\text{ }
\end{eqnarray}%
The (Levi-Civita) Riemann curvature $R_{\left\{ {}\right\} }^{\alpha
}{}_{\beta \mu \nu }$ is only a constituent part of the affine curvature $%
R_{\Gamma }^{\alpha }{}_{\beta \mu \nu }$. Since $R_{\left\{ {}\right\}
}^{\alpha }{}_{\beta \mu \nu }$ depends on the metric $g_{\mu \nu }$, in
terms of $\left\{ _{\beta }{}^{\alpha }{}_{\nu }\right\} $, it is referred
to as the \textsl{curvature of the metric}. On the other hand, since $%
\left\{ _{\mu }{}^{\rho }{}_{\nu }\right\} $ can be rewritten as 
\begin{equation}
\left\{ _{\mu }{}^{\rho }{}_{\nu }\right\} =\eta _{IJ}\eta
^{KL}b_{K}{}^{\rho }b_{L}{}^{\sigma }\left( b^{I}{}_{(\nu }\partial _{|\mu
|}b^{J}{}_{\sigma )}+b^{I}{}_{(\mu }\partial _{|\nu |}b^{J}{}_{\sigma
)}-b^{I}{}_{(\mu }\partial _{|\sigma |}b^{J}{}_{\nu )}\right) ,
\end{equation}%
the curvature $R_{\left\{ {}\right\} }^{\alpha }{}_{\beta \mu \nu }$ can
also be called{\huge \ }the \textsl{curvature of the frame} $\left\{
b_{I}\right\} $.

\section{Gravity as a translation gauge theory}

\subsection{Transformation groups in gravity theories}

Transformation theory is the essence of the method of modern theoretical
physics. In particle physics various transformation groups are directly
related to the classification of particle and interactions between them. In
relativistic physics the Lorentz group $SO(3,1)$ is related to the special
relativity principle, while the diffeomorphism to the general relativity
principle. In Yang-Mills type theories a global transformation group
corresponds to a conservation law. Its localization (gauging) leads to a
interaction, the gauge field which is called the connection in the geometric
language. The conservation current is the source and the charge of the
interaction. The connection represents the potential of the gauge field
while the curvature represents its field strength. These statements
constitute the so-called the gauge principle. It should be noted that the
gauge principle is not applied to Lorentz group. In other words, in
Yang-Mills type theories the Lorentz group does not appear as a gauge group
in spite of its crucial role in the theory. Moreover, in special relativity,
the Lorentz group appear as a gauge group neither. The special relativity
principle means the relativity of the velocity of a reference system. A 
\textsl{global} Lorentz transformation represent a \textsl{uniform} motion
of a reference system relative to another one. A \textsl{local} Lorentz
transformation generally represent a \textsl{accelerating} motion of a
reference system relative to another one. For these reasons a \textsl{global}
Lorentz transformation can be called an \textsl{inertial transformation},
while a\textbf{\ }\textsl{local} Lorentz transformation can be referred%
\textbf{\ }an\textbf{\ }\textsl{accelerating transformation}\textbf{\ }%
\textsl{i.e. a noninertial transformation}. By a localization\textsl{\ }a
Lorentz group is gauged and a corresponding connection is introduced.\textsl{%
\ }However,\textsl{\ this connection does not represent a potential of some
force, but a force itself, the inertial force}. "In other words, there is no
physical gauge field corresponding to the local Lorentz group [29]." On this
account the \textsl{local} Lorentz group is not a gauge group in the
Yang-Mills sense.

As a general coordinate transformation, a diffeomorphism is inevitably
associated with a transformation of the coordinate basis, a $GL(4,\mathbb{R}%
) $ transformation. This is a local (gauged) transformation naturally. The
concept of the noninertial transformation can be \textsl{extended} to local $%
GL(4,\mathbb{R})$ transformations and coordinate transformations. In the
transformation 
\begin{equation*}
b_{K}^{\prime }=\Lambda _{K}{}^{L}b_{L},\ \ \ \Lambda _{K}{}^{L}\in GL(4,%
\mathbb{R})\ 
\end{equation*}%
$\Lambda _{K}{}^{L}$ represents the relative "velocity" of $b_{K}^{\prime }$
to $b_{L}$, while $\partial _{\mu }\Lambda _{K}{}^{L}$ the relative
"acceleration". If $\partial _{\mu }\Lambda _{I}{}^{J}\neq 0$, the
transformation is a \textsl{noninertial transformation}. For the coordinate
transformations, if $\frac{\partial ^{2}x^{\tau }}{\partial x^{\prime \nu
}\partial x^{\prime \mu }}\neq 0$, the transformation $x^{\mu }\rightarrow
x^{\prime \mu }$ is a \textsl{noninertial transformation }too\textsl{\ [42]}.

In these cases the affine connection $\Gamma ^{I}{}_{J\mu }$ or $\Gamma
^{\lambda }{}_{\mu \nu }$ represents a inertial force rather than a
potential. Especially, the flat connections (28, 29, 30, and$\ $31) in
subsection 2.3.1 represent the \textsl{inertial forces} accompanying the
corresponding transformations rather than a potential. \textsl{In this sense
a diffeomorphism or a local }$GL(4,R)$\textsl{\ is not a gauge group in the
Yang-Mills sense.}

The localization of a Lorentz transformation transforms a uniform relative
motion of reference systems to an accelerated one, thereby promotes the
special relativity principle to the general relativity principle. This
promotion involves \textsl{only the inertial force and has nothing to do
with gravity.} Einstein used general coordinate transformations to include
the accelerated relative motion of reference systems and attained the
general relativity principle, then used the equivalence principle to involve
gravity. However, \textsl{as a fictitious force, the inertial force can not
be identified with a real one, the gravitational force}\textbf{.}

In Newton mechanics an inertial force has to be introduced in an accelerated
system in order to hold the second law of motion. In GR a connection is
introduced\ to play the role of an inertial force in a\textsl{\ local }$GL(4,%
\mathbb{R})$ or $SO(1,3)$ transformation.

From the eq.%
\begin{equation*}
e^{I{}}{}_{\mu }:=\partial _{\mu }y^{I}+\omega ^{I}{}_{J\mu
}y^{J}+B^{I{}}{}_{\mu }.\text{ \ \ }
\end{equation*}%
we can see that $\Gamma ^{I}{}_{J\mu }y^{J}$ and $B^{I{}}{}_{\mu }$ have the
same dimension, if $B^{I}{}_{\mu }$ represents a potential, $\Gamma
^{I}{}_{J\mu }$ must represent a field strength, i.e. force. At the same
time, the eq. 
\begin{equation*}
\Gamma ^{\mu }{}_{\alpha \beta }=\left\{ _{\alpha }{}^{\mu }{}_{\beta
}\right\} +K^{\mu }{}_{\alpha \beta }+L^{\mu }{}_{\alpha \beta }
\end{equation*}%
indicates that if $\left\{ _{\alpha }{}^{\mu }{}_{\beta }\right\} $
represents a force, a field strength, according to Einstein, $\Gamma ^{\mu
}{}_{\alpha \beta }$ must represent a force, too. According to it
transformation property, $\Gamma ^{\mu }{}_{\alpha \beta }$ is frame
dependent, therefore, can only represent a \textbf{fictitious force, a
inertial force.}

We conclude that either the Lorentz connection or the affine connection can
only represent a\ force rather than a potential, and therefore, \textsl{%
neither }$SO(1,3)$\textsl{\ nor }$GL(4,R)$\textsl{\ can be considered as a
gauge group in the Yang-Mills sense.} The gauge group of gravity cannot be
anything but the\ translation group\textbf{\ }$T^{4}$.\textbf{\ }

In GR, $K_{\mu \alpha \beta }=0$, $L_{\mu \alpha \beta }=0$, $\Gamma {}_{\mu
\alpha \beta }=\left\{ _{\alpha \mu \beta }\right\} $, $R_{\Gamma }^{\lambda
}{}_{\tau \mu \nu }=R_{\left\{ {}\right\} }^{\lambda }{}_{\tau \mu \nu }$,
the curvature of the affine connection reduces to the curvature of the
Levi-Civita connection. In this case, the connection is not a independent
variable but the derivative of the metric $g_{\mu \nu }$. The metric is the
sole fundamental variable which is the gravitational potential, while the
connection $\Gamma {}_{\mu \alpha \beta }=\left\{ _{\alpha \mu \beta
}\right\} $ is the gravitational field strength. Einstein did not
differentiate the gravity from the inertial force, the coordinate system
from the reference system [42]. It is Einstein who was the first to
associate transformation with interaction, although the general coordinate
transformation is not the gravitational gauge transformation.

It was only after the discovery of the gauge principle and Cartan's tetrad
some years later that it was possible to find the \textsl{genuine
gravitational gauge group}, the translation group $T^{4}$. The gravitational
potential is represented by the connection $B^{I{}}{}_{\mu }$ of $T^{4}$,
the gravitational field strength by its "curvature" $F_{\mu \nu }=\partial
_{\mu }B^{I{}}{}_{\nu }-\partial _{\nu }B^{I{}}{}_{\mu }$.

Both tetrads $h_{\left( \Gamma \right) }^{I}{}_{\mu }=\partial
y^{I}/\partial x^{\mu }+\Gamma ^{I}{}_{J\mu }y^{J}$ and $h_{\left( B\right)
}^{I}{}_{\mu }=\partial y^{I}/\partial x^{\mu }+B^{I}{}_{\mu }$ are
anholonomy, but have different physical meaning. $h_{\left( \Gamma \right)
}^{I}{}_{\mu }$ represents an accelerated reference system, while $h_{\left(
B\right) }^{I}{}_{\mu }$ represents a gravitational field (potential). This
indicates \textsl{the double meaning of the tetrad }$h^{I}{}_{\mu }$ [3]%
\textsl{.}

The formula 
\begin{equation*}
h^{I{}}{}_{\mu }:=\partial _{\mu }y^{I}+\Gamma ^{I}{}_{J\mu
}y^{J}+B^{I{}}{}_{\mu }\text{ \ \ }
\end{equation*}%
indicates that both the translation connection $B^{I{}}{}_{\mu }$ and
affine\ connection $\Gamma ^{I}{}_{J\mu }$ are included in $h^{I{}}{}_{\mu }$
which are the basis of affine transformations. And the torsion $T^{I}{}_{\mu
\nu }$ includes both the curvature $R^{I}{}_{J\mu \nu }$ as well as the
"curvature" $F_{\mu \nu }=\partial _{\mu }B^{I{}}{}_{\nu }-\partial _{\nu
}B^{I{}}{}_{\mu }$ of $B^{I{}}{}_{\mu }$: 
\begin{eqnarray*}
T^{I}{}_{\mu \nu } &=&-R^{I}{}_{J\mu \nu }y^{J}+D_{\nu }B^{I{}}{}_{\mu
}-D_{\mu }B^{I{}}{}_{\nu } \\
&=&-R^{I}{}_{J\mu \nu }y^{J}+\partial _{\nu }B^{I{}}{}_{\mu }-\partial _{\mu
}B^{I{}}{}_{\nu }+\Gamma ^{I}{}_{J\nu }B^{J{}}{}_{\mu }-\Gamma ^{I}{}_{J\mu
}B^{J{}}{}_{\nu }.\ \ \ \ 
\end{eqnarray*}%
One can finds that although $T^{I}{}_{\mu \nu }=0,R^{I}{}_{J\mu \nu }=0,$
the gravitational field strength $F^{I}{}_{\nu \mu }=\partial _{\nu
}B^{I}{}_{\mu }-\partial _{\mu }B^{I}{}_{\nu }=\Gamma ^{I}{}_{J\mu
}B^{J{}}{}_{\nu }-\Gamma ^{I}{}_{J\nu }B^{J{}}{}_{\mu }\neq 0.$ However,
since the connection $\Gamma ^{I}{}_{J\mu }$ is not a tensor, it can be
transformed by a frame transformation such that makes $F^{I}{}_{\nu \mu }=0.$
This is just the so called \textsl{equivalence principle}.

The introduction of diffeomorphism and $SO(1,3)$\ is due to the requirement
of the general relativity principle.\textsl{\ }As in Newton mechanics, the
introduction of accelerated reference systems is necessarily accompanied by
inertial forces. In order to exclude this kind of\textbf{\ fictitious forces
from the dynamics, }in GR the inertial force is identified with gravity
directly, the curvature of the affine connection appears in Einstein's
equation only as the second derivative of the gravitational potential $%
g_{\mu \nu }$.\textbf{\ }In TEGR [1,2] and CGR [4] the connection $\omega
^{I}{}_{J\mu }$ or $\Gamma ^{\lambda }{}_{\mu \nu }$ is directly supposed to
equal zero.

\textsl{In summary, the diffeomorphism or the Lorentz group implements the
relativity principle, whereas the translation group implements the gauge
principle and introduces the gravitational interaction. }

\subsection{\textsl{Two kinds of formulations} of gravitational theories}

\textsl{Gravitational dynamics can be derived from different actions} and
for \textbf{different choices of the dynamic variables}. In order to develop
a gravitational theory, first of all we ought to choose a dynamic variable
to describe gravity. This variable must have a well transformation
character. The potential $B^{I}{}_{\mu }$ is not covariant under the gauge
transformation. Instead, the very suitable variables are the metric $g_{\mu
\nu }$ or the tetrad $h^{I}{}_{\mu }$ which are simple functions of $%
B^{I}{}_{\mu }$ and covariant under all of transformations. This gives rise
to \textsl{two formulations} of gravitational theories, the\textsl{\ metric
formulation }and\textsl{\ the tetrad formulation}. \ In order to derive the
field equation a Lagrangian is needed as the start point. As customary of
modern field theories, the Lagrangian is chosen to be a \textbf{quadratic}
form of the field strength $T^{\lambda }{}_{\mu \nu }$ or $Q^{\lambda
}{}_{\mu \nu }$. At the same time the relativity principle requires that the
Lagrangian must be a scalar, the field equation must be a tensor equation
under the corresponding transformation. A genuine gravity theory should get
rid of the inertial force. Therefore, the affine or Lorentz connection can
not appear in the Lagrangian as an independent dynamic variable. Such that
there is no \textbf{motion} equation of the affine or Lorentz connection,
unless some \textbf{constraint} equations. For this reason, \textsl{the
curvature can appear in the Lagrangian only in a linear form at the most. }%
The Lagrangian of the gravitational field should be a covariant quadratic
form of the first derivative of $g_{\mu \nu }$ or $h^{I}{}_{\mu }$ ($%
e^{I}{}_{\mu }$).

Guided by Ockham's Razor Principle, we seek the simplest Lagrangian.

\subsubsection{Metric formulations}

Einstein's genius was choosing the metric Lagrangian [34]{\small \ } 
\begin{equation}
\mathcal{L}_{E}=\sqrt{-g}g^{\mu \nu }\left( \left\{ _{\mu }{}^{\alpha
}{}_{\beta }\right\} \left\{ _{\alpha }{}^{\beta }{}_{\nu }\right\} -\left\{
_{\mu }{}^{\alpha }{}_{\nu }\right\} \left\{ _{\alpha }{}^{\beta }{}_{\beta
}\right\} \right) ,\ \ \ \ 
\end{equation}%
which is a quadratic form of the field strength $\left\{ _{\mu }{}^{\alpha
}{}_{\nu }\right\} $ of the potential $g_{\mu \nu }$. However, $\mathcal{L}%
_{E}$ is not a scalar. Instead, Hilbert chose $\mathcal{L}_{H}=\sqrt{-g}R$
which is related with $\mathcal{L}_{E}$ by 
\begin{equation}
\mathcal{L}_{H}=-\mathcal{L}_{E}+\partial _{\alpha }B^{\alpha }\text{ \ \ }
\end{equation}%
with%
\begin{equation}
B^{\alpha }=\sqrt{-g}\left( g^{\alpha \mu }\left\{ _{\mu }{}^{\nu }{}_{\nu
}\right\} -g^{\mu \nu }\left\{ _{\mu }{}^{\alpha }{}_{\nu }\right\} \right) ,
\end{equation}%
Dropping the total divergence $\partial _{\alpha }B^{\alpha }$ from $%
\mathcal{L}_{H}$ gives $\mathcal{L}_{E}$. Therefore, they give the same
field equation.

The gravitational Lagrangian $\mathcal{L}_{H}$ gives the field equation%
\begin{equation}
R{}_{\mu \nu }-\frac{1}{2}Rg{}_{\mu \nu }=\left( \delta _{\mu }^{\rho
}\delta _{\nu }^{\sigma }-\frac{1}{2}g{}_{\mu \nu }g^{\rho \sigma }\right)
\nabla _{\lambda }\left\{ _{\rho }{}^{\lambda }{}_{\sigma }\right\}
=T{}_{\mu \nu },
\end{equation}%
where $T{}_{\mu \nu }$ is the energy-momentum of the source matter.
Comparing $\left( 74\right) $ and $(77)$ with the Yang-Mills Lagrangian

\begin{equation*}
\mathcal{L}=\frac{1}{4}F_{I}{}^{\mu \nu }F^{I}{}_{\mu \nu },
\end{equation*}%
and the gauge field equation 
\begin{equation*}
\nabla _{\nu }F_{I}{}^{\mu \nu }=J_{I}{}^{\mu },
\end{equation*}%
one finds that the connection $\left\{ _{\beta }{}^{\alpha }{}_{\gamma
}\right\} $ corresponds to the field strength $F_{I}{}^{\mu \nu }$, while $%
g{}_{\mu \nu }$ corresponds to the Yang- Mills field potential. As it is
known, in electromagnetism and Yang-Mills theory, the connection $%
A^{I}{}_{\mu }$ represents the gauge potential, while the curvature $%
F^{I}{}_{\mu \nu }=\partial _{\mu }A^{I}{}_{\nu }-\partial _{\nu
}A^{I}{}_{\mu }$ represents the force (field strength). The position error
(wrong placement) of the potential and the field strength is the origin of
all the trouble and confusion in GR. It can never been formulated as a
Yang-Mills type gauge theory. We can also say that GR is a "nonnormal" gauge
theory. The "nonnormality" is that the connection $\left\{ _{\beta
}{}^{\alpha }{}_{\gamma }\right\} $ represents the field strength rather
than the potential. However, it is this "nonnormality" that leads to the 
\textsl{equivalent principle}: \emph{the gravitational field strength can be
canceled by a suitable coordinate transformation}. This principle makes GR
differ from any normal gauge theory.

In addition to $\mathcal{L}_{H}$ there is another choice of the
gravitational Lagrangian. We can rewrite Einstein's Lagrangian into a
covariant form by replacing $\partial _{\lambda }g_{\mu \nu }$ with $%
Q_{\lambda \mu \nu }=\nabla _{\lambda }g_{\mu \nu }$. This means that we
suppose $\Gamma ^{\rho }{}_{\mu \nu }\neq \left\{ _{\mu }{}^{\rho }{}_{\nu
}\right\} $, $Q_{\lambda \mu \nu }\neq 0$ and replace 
\begin{equation*}
\left\{ _{\mu }{}^{\rho }{}_{\nu }\right\} =\frac{1}{2}g^{\rho \sigma
}\left( \partial _{\mu }g_{\sigma \nu }+\partial _{\nu }g_{\sigma \mu
}-\partial _{\sigma }g_{\mu \nu }\right) ,
\end{equation*}%
with%
\begin{equation*}
L^{\rho }{}_{\mu \nu }=-\frac{1}{2}g^{\rho \sigma }\left( \nabla _{\mu
}g_{\sigma \nu }+\nabla _{\nu }g_{\sigma \mu }-\nabla _{\sigma }g_{\mu \nu
}\right)
\end{equation*}%
in $\mathcal{L}_{E}$. As thus we get the Lagrangian%
\begin{eqnarray}
\mathcal{L}_{C} &\mathcal{=}&\sqrt{-g}g^{\mu \nu }\left( L^{\alpha
}{}_{\beta \mu }L^{\beta }{}_{\nu \alpha }-L^{\alpha }{}_{\beta \alpha
}L^{\beta }{}_{\mu \nu }\right)  \notag \\
&=&\sqrt{-g}g^{\mu \nu }\left( {}-\frac{1}{4}Q{}_{\alpha \beta \mu
}Q{}^{\alpha \beta \mu }{}+\frac{1}{2}Q_{\alpha \beta }{}{}_{\mu }Q{}^{\beta
\alpha \mu }+\frac{1}{4}Q_{\alpha }{}Q^{\alpha }-\frac{1}{2}Q_{\alpha }{}%
\widetilde{Q}{}^{\alpha }{}{}\right) ,
\end{eqnarray}%
where 
\begin{equation}
Q_{\mu }=Q_{\mu }{}^{\alpha }{}_{\alpha },\widetilde{Q}^{\mu }=Q_{\alpha
}{}^{\mu \alpha }.
\end{equation}%
Then we now compute the \textbf{variations} with respect to the \textbf{%
metric }$g_{\mu \nu }$ and the independent \textbf{connection }$\Gamma
^{\rho }{}_{\mu \nu }$, we arrive at\textbf{\ two} sets of \textbf{field
equations}. Variations with respect to the \textbf{metric }yield equations
that resemble the Einstein field equations:%
\begin{equation}
\partial _{\alpha }\left( \sqrt{-g}\Pi _{\lambda }^{\alpha \mu \nu \rho
\sigma }\left( g\right) L^{\lambda }{}_{\rho \sigma }\right) =\sqrt{-g}%
\left( t^{\mu \nu }+T^{\mu \nu }\right) ,\ \ \ \ 
\end{equation}%
where 
\begin{eqnarray}
\Pi _{\lambda }^{\alpha \mu \nu \rho \sigma }\left( g\right) &=&\frac{1}{2}%
\{\left( g^{\nu \rho }g^{\sigma \mu }+g^{\sigma \nu }g^{\mu \rho }-g^{\rho
\sigma }g^{\nu \mu }\right) \delta _{\lambda }^{\alpha }+\left( g^{\mu \nu
}g^{\rho \alpha }-2g^{\alpha \nu }g^{\rho \mu }\right) \delta _{\lambda
}^{\sigma }  \notag \\
&&+\left( g^{\rho \sigma }g^{\nu \alpha }-g^{\nu \rho }g^{\sigma \alpha
}\right) \delta _{\lambda }^{\mu }+\left( g^{\sigma \alpha }g^{\rho \mu
}-g^{\rho \sigma }g^{\alpha \mu }\right) \delta _{\lambda }^{\nu }\},\text{
\ \ \ \ }
\end{eqnarray}%
\begin{eqnarray}
t^{\mu \nu } &=&\left( \frac{1}{2}g^{\mu \nu }g^{\tau \sigma }-g^{\mu \tau
}g^{\nu \sigma }\right) \left( L^{\rho }{}_{\lambda \tau }L^{\lambda
}{}_{\sigma \rho }-L^{\lambda }{}_{\rho \lambda }L^{\rho }{}_{\tau \sigma
}\right)  \notag \\
&&-g^{\tau \rho }g^{\mu \sigma }\left( L^{\lambda }{}_{\sigma \lambda
}L^{\nu }{}_{\tau \rho }-L^{\nu }{}_{\lambda \tau }L^{\lambda }{}_{\rho
\sigma }+L^{\nu }{}_{\lambda \sigma }L^{\lambda }{}_{\tau \rho }-L^{\lambda
}{}_{\sigma \tau }L^{\nu }{}_{\rho \lambda }\right) ,\text{ \ \ \ }
\end{eqnarray}

\begin{equation}
T^{\mu \nu }=-\frac{2}{\sqrt{-g}}\frac{\delta \mathcal{L}_{\text{m}}}{\delta
g_{\mu \nu }},\text{ \ \ \ \ \ \ \ \ \ \ \ \ }
\end{equation}%
while variations with respect to the \textbf{connection} give a new set of
field equations which determine the connection $\Gamma ^{\rho }{}_{\mu \nu }$%
, 
\begin{equation}
\sqrt{-g}P_{\rho }{}^{\mu \nu }=-\Delta _{\rho }{}^{\mu \nu },
\end{equation}%
where 
\begin{eqnarray}
P^{\rho }{}_{\mu \nu } &=&-\frac{1}{4}Q^{\rho }{}_{\mu \nu }+\frac{1}{2}%
Q_{(\mu }{}^{\rho }{}_{\nu )}+\frac{1}{4}Q^{\rho }g_{\mu \nu }-\frac{1}{4}%
\left( \widetilde{Q}^{\rho }g_{\mu \nu }+\delta _{(\mu }^{\rho }Q_{\nu
)}\right) , \\
\Delta _{\rho }{}^{\mu \nu } &=&-\frac{1}{2}\frac{\delta \mathcal{L}_{\text{m%
}}}{\delta \Gamma ^{\alpha }{}_{\mu \nu }}.
\end{eqnarray}%
The source term $\Delta _{\rho }{}^{\mu \nu }$\ is often referred to as the
hyper-momentum, following a commonly used notation [35].

As expected, the field equations $\left( 84\right) $ are only algebraic
equations of $\Gamma ^{\rho }{}_{\mu \nu }$, which means that they are not
motion equations but constraint equations. Therefor the connection $\Gamma
^{\alpha }{}_{\mu \nu }$ is not a independent dynamic variable. By a gauge
fixing $\Gamma ^{\alpha }{}_{\mu \nu }=0$, we reach a\textbf{\ purified
gravity theory }[36] i.e. CGR, which "\textbf{fundamentally deprives gravity
of any inertial character}\emph{\textbf{"}} and "realises gravitation as a
gauge theory of` translations".[5].

The \textsl{fixed gauge} $\Gamma ^{\alpha }{}_{\mu \nu }=0$ means $L{}_{\mu
\alpha \beta }=-\left\{ _{\alpha \mu \beta }\right\} $, in this case the
Lagrangian $\mathcal{L}_{C}$ goes back to the Einstein's original Lagrangian 
$\mathcal{L}_{E}$, while the Eq. (80) become the Einstein equation (77). 
\textsl{Therefore we have gone back to GR. The so-called coincident GR is
nothing but GR\textbf{.} The distinction is only the geometrical
interpretation yet. }In GR, the curvature of the connection $\left\{ _{\mu
}{}^{\rho }{}_{\nu }\right\} $ is called the spacetime curvature, the
spacetime is a Riemann manifold. In CGR the curvature $R_{\Gamma \beta \mu
\nu }^{\alpha }$ of the connection $\Gamma ^{\rho }{}_{\mu \nu }$ is called
the spacetime curvature, the spacetime is a Minkowski space, although the
curvature $R_{\left\{ {}\right\} \beta \mu \nu }^{\alpha }$ of the
Levi-Civita connection $\left\{ _{\mu }{}^{\rho }{}_{\nu }\right\} $ does
not vanish. The gravitational field is unique, but the geometries describing
it can be different.\ The question is who represents the spacetime? $\Gamma
^{\rho }{}_{\mu \nu }$\ or $\left\{ _{\mu }{}^{\rho }{}_{\nu }\right\} $
(i.e. $g_{\mu \nu }$)?

\subsubsection{Tetrad formulations}

If we choose $B^{I{}}{}_{\mu }$ as the fundamental variable to describe
gravity, the torsion, i.e. the field strength of it can be used to construct
the Lagrangian. As the simplest\ quadratic form of the field strength it is
given by [19] 
\begin{equation}
\mathcal{L}_{T}=\frac{1}{2}e\epsilon _{IJKL}e^{I}\wedge e^{J}\wedge
K^{K}{}_{M}\wedge K^{LM}.\ \ 
\end{equation}%
is $\epsilon _{IJKL}$ the Levi-Civita symbol, $K^{K}{}_{M}$ is the
contortion. Rewriting it in terms of the torsion $T^{\rho }{}_{\mu \nu }$,
we have%
\begin{equation}
\mathcal{L}_{T}\text{ }\mathcal{=}\text{ }e\left( \frac{1}{4}T^{\rho
}{}_{\mu \nu }T_{\rho }{}^{\mu \nu }+\frac{1}{2}T^{\rho }{}_{\mu \nu
}T{}^{\nu \mu }{}_{\rho }-T{}_{\rho \mu }{}^{\rho }T{}^{\nu \mu }{}_{\nu
}\right) \ \ \ 
\end{equation}%
where $e=\det \left( e^{I}{}_{\mu }\right) $.

The variation of $\mathcal{L}_{T}$\ with respect to $B^{I}{}_{\rho }$ gives
the field equation: 
\begin{equation}
\partial _{\sigma }\left( eS_{I}{}^{\sigma \rho }\right) =\frac{1}{2}e\left(
t_{I}{}^{\rho }+\Theta _{I}{}^{\rho }\right) ,\text{ \ \ \ \ \ \ \ \ \ \ \ }
\end{equation}%
where 
\begin{equation}
S^{\rho \mu \nu }=\frac{1}{2}\left( K^{\mu \nu \rho }-g^{\rho \nu }T^{\sigma
\mu }{}_{\sigma }+g^{\rho \mu }T^{\sigma \nu }{}_{\sigma }\right) ,\text{ \ }
\end{equation}%
and%
\begin{equation}
et_{I}{}^{\rho }=2ee_{I}{}^{\lambda }S_{\mu }{}^{\nu \rho }T^{\mu }{}_{\nu
\lambda }-e_{I}{}^{\rho }\mathcal{L_{T}\ \ \ }
\end{equation}%
is the energy-momentum tensor of gravitation and $e\Theta _{\lambda
}{}^{\rho }$ is the source energy-momentum tensor.

Since 
\begin{equation*}
e^{I{}}{}_{\mu }=\partial _{\mu }y^{I}+\omega ^{I}{}_{J\mu
}y^{J}+B^{I{}}{}_{\mu },
\end{equation*}%
includes $\omega ^{I}{}_{J\mu }$, the variation of $\mathcal{L}_{T}$\ with
respect to $\omega ^{I}{}_{J\mu }$ gives the field equation:

\begin{eqnarray}
&&\frac{\partial \left( e\Pi _{IJ}{}^{\mu \nu \rho \sigma }\right) }{%
\partial \omega ^{M}{}_{N\lambda }}T^{I}{}_{\mu \nu }T^{J}{}_{\rho \sigma } 
\notag \\
&&+e\left\{ \Pi _{IJ}{}^{\mu \nu \rho \sigma }+\Pi _{JI}{}^{\rho \sigma \mu
\nu }\right\} \left\{ -\delta _{M}^{I}\delta _{K}^{N}\left( \delta _{\mu
}^{\tau }\delta _{\nu }^{\lambda }-\delta _{\nu }^{\tau }\delta _{\mu
}^{\lambda }\right) y^{K}+\delta _{M}^{I}\delta _{K}^{N}\left( \delta _{\nu
}^{\lambda }B^{K}{}_{\mu }-\delta _{\mu }^{\lambda }B^{K}{}_{\nu }\right)
\right\} T^{J}{}_{\rho \sigma }  \notag \\
&&+\ \partial _{\tau }\left( e\left\{ \Pi _{IJ}{}^{\mu \nu \rho \sigma }+\Pi
_{JI}{}^{\rho \sigma \mu \nu }\right\} \delta _{M}^{I}\left( \delta _{\mu
}^{\tau }\delta _{\nu }^{\lambda }-\delta _{\nu }^{\tau }\delta _{\mu
}^{\lambda }\right) y^{N}T^{J}{}_{\rho \sigma }\right)  \notag \\
&=&\Delta _{M}{}^{N\lambda },\ \ \ \ \ \ \ \ \ \ \ \ \ \ \ \ \ \ \ \ \ \ \ \
\ \ \ \ \ \ \ \ \ \ \ \ \ \ \ \ \ \ \ \ \ \ \ \ \ \ \ \ \ \ \ 
\end{eqnarray}%
where 
\begin{equation}
\Pi _{IJ}{}^{\mu \nu \rho \sigma }\left( e\right) =\left( \frac{1}{4}\eta
_{IJ}{}g^{\sigma \nu }+\frac{1}{2}e_{I}{}^{\sigma }e_{J}{}^{\nu
}-e_{I}{}^{\nu }e_{J}{}^{\sigma }\right) g^{\rho \mu },\text{\ \ \ }
\end{equation}%
\begin{equation}
\Delta _{M}{}^{N\lambda }=-\frac{\delta \mathcal{L}_{\text{m}}}{\delta
\omega ^{M}{}_{N\lambda }}.
\end{equation}

The equation%
\begin{equation*}
T^{I}{}_{\mu \nu }=-R^{I}{}_{J\mu \nu }y^{J}+D_{\nu }B^{I{}}{}_{\mu }-D_{\mu
}B^{I{}}{}_{\nu }.\ \ \ \ 
\end{equation*}%
indicates that (92) is a second order differential equation of $\omega
^{I}{}_{J\mu }$, i.e. a motion equation of $\omega ^{M}{}_{N\lambda }$.
However, in the case 
\begin{equation*}
R^{I}{}_{J\mu \nu }=0,
\end{equation*}%
(92) degenerates to a fist order differential equation, which means that it
is no longer a motion equation but a \textsl{constraint} equation on $\omega
^{I}{}_{J\mu }$ and $\omega ^{I}{}_{J\mu }$ is no longer a dynamic variable
either. In this case, $\mathcal{L}_{T}$ is exactly the Hilbert Lagrangian up
to a divergences, and (89) is equivalent to Einstein's equation. The theory
becomes a\textsl{\ teleparallel gravity theory}. In the case $\omega
^{I}{}_{J\mu }=0$, we obtain TEGR which was proposed as an alternative
formulation of general relativity by Einstein [37].

When $\omega ^{I}{}_{J\mu }=0$, we have

\begin{equation*}
e^{I{}}{}_{\mu }=e^{\left( 0\right) I{}}{}_{\mu }=\partial _{\mu
}y^{I}+B^{I{}}{}_{\mu },
\end{equation*}%
and 
\begin{equation*}
T^{I}{}_{\mu \nu }=\partial _{\nu }e^{\left( 0\right) I{}}{}_{\mu }-\partial
_{\mu }e^{\left( 0\right) I{}}{}_{\nu }=\partial _{\nu }B^{I{}}{}_{\mu
}-\partial _{\mu }B^{I{}}{}_{\nu }=-F^{I}{}_{\mu \nu }.
\end{equation*}%
The theory becomes the so-called\ pure tetrad\ teleparallel gravity theory.
The gravitational field strength $F^{I}{}_{\mu \nu }$\ is the anholonomy of
the tetrad $e^{\left( 0\right) I{}}{}_{\mu }{}$.

There is another choice $B^{I{}}{}_{\mu }=0$\ but $\omega ^{I}{}_{J\mu }\neq
0$ for the Lagrangian $\mathcal{L}_{T}$, which means $T^{I}{}_{\mu \nu
}=-R^{I}{}_{J\mu \nu }y^{J}$. In this case,$\ (92)\ $becomes$\ $the field
equation of $\omega ^{I}{}_{J\mu }y^{J}$, which has the same form of the
equation (89) of $B^{I{}}{}_{\mu }$ with $\omega ^{I}{}_{J\mu }=0$, but
obeys the different transformation law [39]. In this case $\omega
^{I}{}_{J\mu }y^{J}$ amounts to a "potential" of the force $\omega
^{I}{}_{J\mu }$, while $T^{I}{}_{\mu \nu }$ to the strength. In this sense,
the inertial force $\omega ^{I}{}_{J\mu }$ can be identified with the
gravitational force $T^{I}{}_{\mu \nu }=-F^{I}{}_{\mu \nu }$, as expected by
Einstein. It is in this case that $\omega ^{I}{}_{J\mu }$ can be expressed
as a Lorentz gauge potential [38] (but is not the gravitational potential
which is $\omega ^{I}{}_{J\mu }y^{J}$ rather than $\omega ^{I}{}_{J\mu }$!
In other words, $\omega ^{I}{}_{J\mu }$ corresponds to rather than equals to
the gravitational potential! Note: the local Lorentz rotation $\delta
y^{I}=\varepsilon ^{I}{}_{J}\left( x\right) y^{J}$\ is equal to a local
translation $\delta y^{I}=\varepsilon ^{I}{}\left( x\right) $) and the
curvature can be identified with the gravitational field strength $%
T^{I}{}_{\mu \nu }=-R^{I}{}_{J\mu \nu }y^{J}$. However, this case is no
longer belong to the teleparallel theories but Lorentz gauge theories [39],
since the Lagrangian $L_{T}$ becomes 
\begin{equation}
\mathcal{L}_{L}\mathcal{=}e\left( \frac{1}{4}\eta _{IJ}g^{\mu \lambda
}g^{\nu \tau }+\frac{1}{2}e_{I}{}^{\tau }e_{J}{}^{\nu }g^{\mu \lambda
}-e_{I}{}^{\nu }e_{J}{}^{\tau }g^{\mu \lambda }\right) R^{I}{}_{K\mu \nu
}R^{J}{}_{L\lambda \tau }y^{K}y^{L}.
\end{equation}%
Then General Relativity is \textbf{not recovered, }instead yielding a
modified theory where, for example, gravitational waves do not travel at the
speed of light [40]\textbf{. }In addition, in the Lorentz gauge theory there
exists \textsl{only the angular momentum (AM) conservation} with respect to
the Lorentz symmetry, but the energy-momentum (EM) conservation is absent
[41]. In other words, the energy-momentum is not the source of the field,
the theory is failed to be a gravity theory. Even in the Lorentz gauge
theory, the angular momentum conservation corresponds to only a subgroup $%
SO(3)$, whereas the conservation current corresponding to Lorentz
transformation is absent.

\subsubsection{Remarks}

In GR $T^{\rho }{}_{\mu \nu }=0$, $Q^{\rho }{}_{\mu \nu }=0$, the connection 
$\Gamma ^{\rho }{}_{\mu \nu }=\left\{ _{\mu }{}^{\rho }{}_{\mu }\right\} $
is the derivative of the metric $g_{\mu \nu }$ and not a independent
variable. Only the metric $g_{\mu \nu }$ is the dynamic variable describing
gravity.

In TEGR $\omega ^{I}{}_{J\mu }=0,$ the connection $\Gamma ^{\rho }{}_{\mu
\nu }=e_{I}{}^{\rho }e^{I}{}_{\nu ,\mu }$ is the derivative of the tetrad $%
e^{I}{}_{\mu }$ and not a independent variable. Only the tetrad $%
e^{I}{}_{\mu }$ is the dynamic variable describing gravity.

In CGR $\Gamma ^{\rho }{}_{\mu \nu }=0,$ the connection $\Gamma ^{I}{}_{J\mu
}=e^{I}{}_{\nu }\partial _{\mu }e_{J}{}^{\nu }$ is the derivative of the
cotetrad $e_{J}{}^{\nu }$ and not a independent variable. Only the cotetrad $%
e_{J}{}^{\nu }$ is the dynamic variable describing gravity.

Since $g_{\mu \nu }=\eta _{IJ}e^{I}{}_{\mu }e^{J}{}_{\nu },$we can say that
in all these three cases only the\ (co)tetrad is the dynamic variable
describing gravity. Recalling $e^{I{}}{}_{\mu }=\partial _{\mu }y^{I}+\omega
^{I}{}_{J\mu }y^{J}+B^{I{}}{}_{\mu },$ we can say that only the translation
potential $B^{I{}}{}_{\mu }$ is the genuine dynamic variable describing
gravity.

At the same time, about the translation gauge potential, there are some
divergences, besides $B^{I{}}{}_{\mu }$, the tetrad $e^{I}{}_{\mu }$ is also
referred as the translation gauge potential [11]. The tangent space
coordinate $y^{J}$ is also referred as a \textsl{coordinate-scalar} [10] and
a dynamic field [15].

\subsection{A concise historical review}

\subsubsection{Einstein's approach}

In the derivation of general relativity theory (GR) [42], Einstein started
in flat Minkowski space, went over to accelerated frames, and applied
subsequently the equivalence principle. The acceleration was described by
using curvilinear coordinates. Hereby he had to relax the rigidity of the
Minkowski geometry ending up with the Riemannian geometry of GR. In fact,
Einstein had gauged the group $GL(4,R)$ but not the group $SO(1,3)$.
Einstein took the metric $g_{\mu \nu }$\ as the gravitational potential and
the Levi-Civita \ connection $\left\{ _{\mu }{}^{\alpha }{}_{\nu }\right\} $%
\ as the field strength. Einstein did not distinguish coordinate system from
reference system, initially. If he was aware of identifying the reference
system as a Lorentz frame, he might use\ a local Lorentz transformation in a
flat Minkowski geometry to attain an accelerated reference system. In this
case, however, appears only the inertial force, no gravity. In order to
include gravity in a flat geometry, teleparallelism and torsion must be
introduced.

Teleparallel theories of gravity were originally proposed by Einstein in the
late 1920s [43--45]. In 1928 [43], he searched for a more general
geometrical framework than the Riemannian geometry to unify gravity and
electromagnetism. He considered replacing the metric tensor by tetrads $%
h^{I}{}_{\mu }$, which are related to the metric tensor through $g_{\mu \nu
}=\eta _{IJ}h^{I}{}_{\mu }h^{J}{}_{\nu }$. He introduced a new geometry by
defining a new covariant derivative $\nabla _{\mu }$ which satisfies the
condition 
\begin{equation}
\nabla _{\mu }h_{I}{}^{\mu }=0.\text{ \ \ \ \ \ \ \ \ }
\end{equation}%
This corresponds a new connection 
\begin{equation}
\Gamma ^{\rho }{}_{\mu \nu }=h_{I}{}^{\rho }\partial _{\nu }h^{I}{}_{\mu },%
\text{ \ \ \ }
\end{equation}%
with a zero curvature. It is just the flat connection (30) found by Weitzenb%
\"{o}ck first [46, 47]

In 1929 [45], Einstein considered all torsional invariants consisting of the
torsion $T^{\rho }{}_{\mu \nu }=\Gamma ^{\rho }{}_{\mu \nu }-\Gamma ^{\rho
}{}_{\nu \mu }$ and showed that there is an unique combination of three
invariants 
\begin{equation}
\mathcal{L}_{\text{E}}=\frac{h}{2\kappa }\left( \frac{1}{4}T^{\rho }{}_{\mu
\nu }{}T{}_{\rho }{}^{\mu \nu }+\frac{1}{2}T^{\rho }{}_{\mu \nu }{}T{}^{\nu
\mu }{}_{\rho }-T^{\nu \mu }{}_{\nu }{}T{}^{\nu }{}_{\mu \nu }\right) ,\text{
\ \ }
\end{equation}%
which is just the \textsl{Lagrangian of} the teleparallel equivalent of
general relativity (\textsl{TEGR}). Apparently, Einstein had been close to
completing this theory. However, his intention was to unify gravity and
electromagnetism rather than develop a new gravitational theory. After all,
GR was invented with great success by himself and there is no need to
modified in his time. "This naturally leads us to the question of how would
modern gravitational research look like if Einstein had continued to work on
the concepts surrounding absolute teleparallelism, or if he had formulated
GR in terms of torsion from the beginning (namely, TEGR) [48]?" It is
Einstein's pursuit of the unified field theory and the finding of gauge
field theories that spawns various gravitational gauge theories and leads to
the translation gauge theory of gravity finally.

In 1960s and 1970s it became clear that Einstein teleparallelism can be a
successful theory as long as we consider it as a theory of gravity only.%
{\small \ }\textsl{It is Einstein who founded not only GR but also TEGR}.

\subsubsection{Gauge theories of gravity}

Shortly after Yang-Mills theory was proposed, Utiyama applied\ the gauge
idea to gravity [23] by using the Lorentz group $SO(1,3)$ as a gauge group.
However, It is unsuccessful, since the current coupling to the Lorentz group
is the angular momentum current rather than the energy-momentum current. The
energy-momentum couples to the translation group $T^{4}$.

Minkowski space has the Poincare group, the semi-direct product of the
translation group $T^{4}$ and the Lorentz group $SO(1,3)$, as its symmetry
group. In other words, Minkowski space is invariant under rigid (`global')
Poincare transformations. The corresponding field-theoretical currents are
the material energy-momentum and spin angular momentum currents,
respectively. Therefore, one may attempt to extend the gauging of the
translations to the gauging of full Poincare group thereby also including
the conservation law of the angular momentum current. The gauging, that is,
the\ localization of the Poincare group, yields the Poincare gauge theory of
gravity (PG). Whereupon the Einstein-Cartan theory [49,50,51] was proposed
as a Poincare gauge theory of gravity.

In the general form of the Poincare gauge framework there exists 4 \textbf{%
translational} gauge potentials $e_{\mu }{}^{I}$ ("gravitons",
\textquotedblleft \textsl{weak Einstein gravity}\textquotedblright ) and 6 
\textbf{Lorentz} gauge potentials $\Gamma _{\mu }{}^{IJ}$ ("rotons",
\textquotedblleft \textsl{strong Yang-Mills gravity}\textquotedblright )
which \textsl{couple} to the \textbf{momentum current} and the \textbf{spin
current} of matter, respectively [52].

However, the\textbf{\ rotons\ have not been observed} so far, one has to try
to \textbf{suppress} them in the theory. There exist two possibilities of
getting rid of the rotons: The Einstein-Cartan-Sciama- Kibble (ECSK)-theory
of gravity or Teleparallel gravity [11,53].

The ECSK field equations are equivalent to the Einstein equations. For
matter without spin the torsion vanishes, and the ECSK theory coincides with
GR.

\ If the spin of matter is suppressed, a In\"{o}n\"{u}-Wigner type\ group
contraction of the PG leads to a translation gauge theory i.e. the
teleparallelism theory. Teleparallelism has a number of unexpected and
somewhat strange features. After all, the vanishing of the curvature, which
is the defining characteristics of teleparallelism theory, is hard to digest
from a purely GR point of view. However, from the point of view of PG, this
is self-evident, since the curvature is the gauge field strength of the
Lorentz group. The suppression of the material spin, in turn, suppresses the
Lorentz group as gauge group. For this reason it may be considered that\
teleparallelism can only be really understood in the context of PG. It is
not comprehensible as a stand-alone theory [12].

By Noether's theorem, energy-momentum conservation is induced by
translational invariance of the Lagrangian of an isolated system. Following
the idea of the Yang-Mills gauge theory, we can recognize that the
translation group $T^{4}$ is the gauge group of ordinary gravity. This point
was already made in the beginning of 1960s by many well-known physicists
[54-57].

Until after the 1970s, the early systematic descriptions of a translational
gauge theory of gravity begin to appear [58, 59]. Late in the last century
this theory achieved unprecedented development, a plethora of papers have
been published of which reviews see [2,3,22,60] for example.

Nevertheless, even after almost 100 years since the initial formulation of
Einstein, debates persist regarding the very foundations of teleparallel
theories of gravity. They primarily revolves around the\ nature of
connections, since these connections can be transformed to zero and hence
effectively\ eliminated from the theory. The main question then concerns
whether to consider these connections as a fundamental variable of the
theory. It is declared even that this teleparallelism scheme cannot be
directly compared with nature [61].

Until the end of the last century, the situation is gradually becoming
clearer with the publication of an article [4]. People have come to realize
that gravity as described by teleparallel theory has noting to do with\
either curvature or torsion. As a translation gauge theory, gravity can be
formulated with neither the Lorentz connection nor the affine connection
which represent only a\ purely fictitious force [5,9,27,29,62] i.e. an
inertial force rather than a genuine force. The\ genuine gravitational force
is represented by translation gauge field strength. Formulating gravity as a
translation gauge theory has been proposed as the first\ principle of motion
[63]. This is the most crucial Lesson teaching us by the geometrical trinity
of gravity [1], which leads to a series of profound changes of fundamental
concepts in gravity theories.

\section{Discussion}

\subsection{Covariantisation}

In recent years the invariance of teleparallel gravity theories becomes a
hot topic. In the current literature, there are many discussions on the
covariant approach [16,17,21,22,64,65] to (modified) teleparallel theories.
The conclusions range all the way from total denial [18] through an
undecided stance [8] to taking it as the only consistent version of the
theory [20].

In fact since%
\begin{equation*}
e^{I{}}{}_{\mu }=\partial _{\mu }y^{I}+\omega ^{I}{}_{J\mu
}y^{J}+B^{I{}}{}_{\mu }
\end{equation*}%
the Lagrangian%
\begin{equation*}
\mathcal{L}_{T}\text{ }\mathcal{=}\text{ }e\left( \frac{1}{4}T^{\rho
}{}_{\mu \nu }T_{\rho }{}^{\mu \nu }+\frac{1}{2}T^{\rho }{}_{\mu \nu
}T{}^{\nu \mu }{}_{\rho }-T{}_{\rho \mu }{}^{\rho }T{}^{\nu \mu }{}_{\nu
}\right) \ \ \ 
\end{equation*}%
is invariant under both local Lorentz and coordinate transformations and
then the theory is covariant. However, in TEGR, a condition $\omega
^{I}{}_{J\mu }=0$ is proposed, while in CGR a condition $\Gamma ^{\alpha
}{}_{\mu \nu }=0$ is proposed. In these cases, the covariance is breaked,
since, recalling 2.1.1, the introduction of the connections $\omega
^{I}{}_{J\mu }$ and $\Gamma ^{\alpha }{}_{\mu \nu }$ is to render the
corresponding covariance. Therefore, in order to restore the covariance they
must be \textsl{reintroduced} by corresponding transformation. In order to
restore the local Lorentz covariance a connection

\begin{equation*}
\omega ^{K}{}_{I\mu }=\left( \Lambda ^{-1}\right) _{J}{}^{K}\partial _{\mu
}\Lambda _{I}{}^{J}.\ \ \ 
\end{equation*}%
must be reintroduced by a Lorentz transformation $\Lambda _{I}{}^{J}\left(
x\right) $ in TEGR [60]. In order to restore the coordinate covariance a
connection 
\begin{equation*}
\Gamma ^{\lambda }{}_{\mu \nu }=\frac{\partial x^{\prime \lambda }}{\partial
x^{\tau }}\frac{\partial ^{2}x^{\tau }}{\partial x^{\prime \mu }\partial
x^{\prime \nu }}\ .\ 
\end{equation*}%
must be reintroduced by a coordinate transformation $x^{\mu }\rightarrow
x^{\prime \mu }\left( x\right) $ in CGR [5,66].

It should be noted that this procedure of\ covariantisation in principle
should not change the physical content of the theory, since it just
introduces a inertial force using a noninertial transformation. The
covariantized\ formulations\ is fully equivalent to the initial one [67,68].
However, this fact\ has not been understood well by the community [21]. It
has generated confusion in modified teleparallel theories and created the
common belief that the covariant modified teleparallel gravity cures all the
problems associated with breaking the Lorentz invariance, which is however
not true [69].

Covariantisation procedures cannot work the same way as in TEGR when
modified teleparallel gravities are generically depend on the tetrad and the
torsion. Since the dependence on the Lorentz connection in generalized
models cannot be reduced to a surface term, the variation with respect to
the Lorentz connection does\ produce non-trivial equation of motion [60].

However, it is proved that in any\ covariant teleparallel model this
equation is redundant, it coincides with the antisymmetric part of the
equation of motion for the tetrad field [17,19]. Therefore, any variation of
the action with respect to the flat Lorentz connection can be precisely
compensated by an appropriate Lorentz transformation of the tetrad. However,
the latter is a legitimate variation of the tetrad field which by itself
must keep the action stationary, as is independently ensured by the
antisymmetric part of the tetrad equation of motion [70, 71].

Therefore, the local Lorentz invariance is not natural at all for the
modified teleparallel theories, the pure tetrad formulation i.e. TEGR is
more fundamentally justified [68]. It is better suited to the geometrical
meaning of the Weitzenb\"{o}ck connection. At the same time, it is of course
possible to rewrite the formulation in terms of an arbitrary basis in the
tangent space, thus getting the covariant form; possible but by no means
necessary [21].

The action of STEGR is identical to the historical $\Gamma \Gamma $ action
of Einstein. It lacks proper\ covariance under diffeomorphisms. Therefore,
the simple generalizations of the STEGR action break diffeomorphisms
symmetry, the same way as local Lorentz invariance gets\ broken in modified
metric teleparallel models [17,72]. We can covariantise it by introducing a\
connection $\Gamma ^{\lambda }{}_{\mu \nu }=\frac{\partial x^{\prime \lambda
}}{\partial x^{\tau }}\frac{\partial ^{2}x^{\tau }}{\partial x^{\prime \mu
}\partial x^{\prime \nu }}$ as in the case of CGR [66].

In numerous papers the covariant symmetric teleparallel theories are taken
as just a fully equivalent rewriting of the coincident-gauge ones, without
rigorously proving this statement. At the same time, some other works do
treat modified covariant STEGR and modified $\Gamma \Gamma $ gravity as\
different models [73, 74]. There is no consensus in the literature.
Consequently, this raise the\ pertinent question whether it is worthwhile
pursuing the covariant approach [66].

These processes of covariantisation can be understood as a Stuckelberg
procedure which in principle should not change the physical content of the
theory. The Stuckelberg trick is very artificial in its nature, and
therefore should not be taken as a\ must. It is asserted that the
covariantized\ formulations\ is fully equivalent to the original one but is
not necessary either [68,21].

\subsection{\textbf{General Covariance Principle}}

Einstein's GR includes two tasks: to\ extend special relativity principle to
involve accelerated reference system and\ to develop a relativistic
gravitational theory. It starts from two principles, the general relativity
principle and the equivalence principle. They are two different principles.
The only link between them is that both involve an accelerated reference
system.

The general relativity principle can be narrated as such that equations
formulating physical laws are covariant under any transformations [42]. It
is also referred as the General Covariance Principle and has nothing to do
with gravity.

However, in this case the general covariance alone is empty of any physical
content. As early as 1917, Kretschmann criticized it and asserted that any
arbitrary theory can be put into a generally covariant form, without
changing its physical content [75].

In this sense "the notion of general relativity does\ not in fact introduce
any postNewtonian physics;\ it simply deals with coordinate transformations.
Such a formalism may have some convenience, but physically it is wholly
irrelevant." Therefore, "general relativity is a physically meaningless
phrase that can only be viewed as a historical memento of a curious
philosophical observation.\textquotedblright \lbrack 76]

The only meaning of the general relativity principle is the realization of
the relativity of acceleration [77], an issue seldom recognized in the
context of general relativity (GR) [78--80].

There are only the two tenable positions [81]: either all frames in GR are
inertial (a viewpoint advocated by e.g. Anderson and Norton) or no frame is
inertial (as e.g. Synge and Fock had argued) [82]. A more reasonable
suggestion is using the concept of a inertial or non-inertial transformation
instead of the concept of a inertial or non-inertial frame. Only relative
acceleration of bodies is supposed to\ be observable. The acceleration of a
frame is relative, while the acceleration of a transformation is absolute.

For some time, the idea of a reference frame was ambiguous, it was mixed
with the idea of a coordinate system. Nowadays, there is a\ clear
distinction between them. Coordinates are used to assign four numbers to
events on a spacetime. In the mathematical language a coordinate system is a
chart on the spacetime manifold, which is meaningless physically. On the
other hand\ a reference system can be considered as a simplified geometric
representation of the quantized measuring instruments. It determine the
metric and various physical objects. In general, the base of a coordinate
system is holonomic, while a reference frame is anholonomic.

\subsection{The meaning of EP}

Galileo's experiment on the Leaning Tower of Pizza indicates that "all
bodies fall the same way in a gravitational field". This result is elevated
to the Equivalence Principle\ as one of the foundations of GR [83].

This principle is not new in Newton's theory of gravitation, new was
Einstein's expression\ [84], where Einstein's elevators displaces Galileo's
falling objects, a gravitational field is assumed to be equivalent to a
corresponding acceleration of the reference system.

Since then this principle has been formulated in various different
expressions. According to the popular expression in the literature, this
principle means the equivalence of the inertial force with the gravity.
However, the inertial force \emph{has no source and is generated by a
coordinate transformation and then can not be equivalent to the real
gravitational fields. }To be considered as \ equivalence is only due to the 
\textbf{universal character} of both them in common. It is the universality
that as a hint was used by Einstein towards general relativity.

Effects equally felt by all bodies were known since long. They are the 
\textit{inertial} effects, which show up in non-inertial frames, for
examples, the centrifugal and the Coriolis forces. On the other hand, the
universality of free fall is the most fundamental characteristic of the
gravitational interaction [33]. However, as mentioned above, inertial force
and gravity are essentially two different kinds of force. And moreover, they
follow different transformation rules, the inertial force is a connection,
while the gravitational force as a translation gauge field strength is a
tensor. It is because of this difference of transformation rules that makes
them can be equivalent or canceled each other by some transformation, which
is the essence of the equivalence principle. It is important to remark that 
\textbf{only} an interaction presenting the property of \textbf{universality}
can be described by a \textbf{geometrization} of spacetime. Since both the
inertial and the gravitational force are universal they can be \textbf{%
geometrized }and canceled each other.

According to%
\begin{equation*}
\Gamma ^{\mu }{}_{\alpha \beta }=\left\{ _{\alpha }{}^{\mu }{}_{\beta
}\right\} +K^{\mu }{}_{\alpha \beta }+L^{\mu }{}_{\alpha \beta }
\end{equation*}%
the affine connection $\Gamma ^{\mu }{}_{\alpha \beta }$ is not equal to the
Levi-Civita connection $\left\{ _{\alpha }{}^{\mu }{}_{\beta }\right\} $. In
GR, therefor, the torsion $T^{\mu }{}_{\alpha \beta }=\Gamma ^{\mu
}{}_{\alpha \beta }-\Gamma ^{\mu }{}_{\beta \alpha }$ can not vanish in
spite of $\left\{ _{\alpha }{}^{\mu }{}_{\beta }\right\} $ is torsionfree.
Since $Q^{\mu }{}_{\alpha \beta }=0$, i.e. $L^{\mu }{}_{\alpha \beta }=0$, $%
\left\{ _{\alpha }{}^{\mu }{}_{\beta }\right\} $\ contains two parts: 
\begin{equation*}
\left\{ _{\alpha }{}^{\mu }{}_{\beta }\right\} =\Gamma ^{\mu }{}_{\alpha
\beta }-K^{\mu }{}_{\alpha \beta }.
\end{equation*}%
which are the inertial force $\Gamma ^{\mu }{}_{\alpha \beta }$ and the
gravitational force $K^{\mu }{}_{\alpha \beta }$. However, since they follow
different transformation rules, an appropriate transformation can leads to $%
\left\{ _{\alpha }{}^{\mu }{}_{\beta }\right\} =0$, which is just the
equivalent principle in GR.

According to%
\begin{equation*}
T^{I}{}_{\mu \nu }=-R^{I}{}_{J\mu \nu }y^{J}+D_{\nu }B^{I{}}{}_{\mu }-D_{\mu
}B^{I{}}{}_{\nu }\ \ \ \ 
\end{equation*}%
and%
\begin{equation*}
T^{I}{}_{\mu \nu }=D_{\nu }h^{I}{}_{\mu }-D_{\mu }h^{I}{}_{\nu }=\partial
_{\nu }h^{I}{}_{\mu }-\partial _{\mu }h^{I}{}_{\nu }+\omega ^{I}{}_{J\nu
}h^{J}{}_{\mu }-\omega ^{I}{}_{J\mu }h^{J}{}_{\nu }
\end{equation*}%
an appropriate Lorentz transformation satisfying%
\begin{equation*}
\omega ^{I}{}_{J\mu }h^{J}{}_{\nu }-\omega ^{I}{}_{J\nu }h^{J}{}_{\mu
}=\partial _{\nu }h^{I}{}_{\mu }-\partial _{\mu }h^{I}{}_{\nu }
\end{equation*}%
can leads to%
\begin{equation*}
T^{I}{}_{\mu \nu }=0.
\end{equation*}%
This means that the inertial force $\omega ^{I}{}_{J\mu }$ cancels the
gravitational force $F^{I}{}_{\mu \nu }=\partial _{\mu }h^{I}{}_{\nu
}-\partial _{\nu }h^{I}{}_{\mu }$, which is just the equivalent principle in
TEGR.

In CGR, 
\begin{eqnarray*}
Q_{\mu \alpha \beta } &\equiv &\nabla _{\mu }g_{\alpha \beta }=\partial
_{\mu }g_{\alpha \beta }-\Gamma ^{\rho }{}_{\alpha \mu }g_{\rho \beta
}-\Gamma ^{\rho }{}_{\beta \mu }g_{\alpha \rho } \\
&=&\partial _{\mu }g_{\alpha \beta }-\Gamma {}_{\beta \alpha \mu }-\Gamma
{}_{\alpha \beta \mu }.
\end{eqnarray*}%
indicates that an appropriate coordinate transformation satisfying%
\begin{equation*}
\partial _{\mu }g_{\alpha \beta }=\Gamma {}_{\beta \alpha \mu }+\Gamma
{}_{\alpha \beta \mu }
\end{equation*}%
makes%
\begin{equation*}
\nabla _{\mu }g_{\alpha \beta }=0,
\end{equation*}%
which means that the inertial force $\Gamma {}_{\alpha \beta \mu }$ cancels
the gravitational force $\partial _{\mu }g_{\alpha \beta }$. This is the
equivalent principle in CGR.

In Newton mechanics the inertial force appears when a reference system is
accelerated, in other words, it appears in an accelerating transformation of
reference systems. In special relativity, the inertial force appears in a
local Lorentz transformation as a Lorentz connection [29]. Therefore, the
inertial force appears in any relativistic theories, for example, Yang-Mills
gauge theories. However, apart from gravity, other three interactions do not
have the \textsl{universality}\textbf{\ }and\textbf{\ }then have nothing to
do with inertial. There is no equivalence principle for them. In fact,
except the common \textsl{universality}\textbf{, }gravity itself has also
nothing to do with inertial force. One can developed a relativistic gravity
theory with neither the inertial force nor the equivalence principle as a
field theory.

As early as in 1960, John Synge confesses [85]: \textquotedblleft . . . I
have never been able to understand this Principle.\textquotedblright\ \
\textquotedblleft Does it mean that the effects of a gravitational field are
indistinguishable from the effects of an observer's acceleration? If so, it
is false. ... The Principle of Equivalence performed the essential office of
midwife at the birth of general relativity, but, as Einstein remarked, the
infant would never have got beyond its long-clothes had it not been for
Minkowski's concept. I suggest that the midwife be now buried with
appropriate honours and the facts of absolute space-time
faced.\textquotedblright

The theoretical relevance of the equivalence principle is mere the
indication of the geometric nature of gravity. The modern view was stated by
Bondi in 1979 [76]: \textquotedblleft From this point of view, Einstein's
elevators have nothing to do with gravitation; they simply analyze inertia
in a perfectly Newtonian way.\textquotedblright

\section{\textbf{Conclusions}}

Gravitational theory stems from Galileo's Leaning Tower of Pisa experiment
which reveals the two features of gravity: (a) the source and charge of
gravity is the mass; (b) gravity is universal. In fact (a) is the reason of
(b), as shown by Newton's law of gravity. In modern gravity theories (a) is
promoted as the statement that the conservation current of gravity is the
energy-momentum, which directly leads to that gravity is a translation gauge
field. The feature (b) leads to the geometrization of gravity and the
equivalence principle. The equivalence principle is based on the
universality of \textbf{both} gravity as well as inertial force.

In Newton theory in order to render the second law of motion \textsl{%
invariant} under \textsl{a accelerated} transformation an inertial force $%
F=ma$ must be introduced, in GR in order to render all relativistic theories 
\textsl{invariant} under \textsl{local Lorentz} transformations an inertial
force $\omega ^{K}{}_{I\mu }=\left( \Lambda ^{-1}\right) _{J}{}^{K}\partial
_{\mu }\Lambda _{I}{}^{J}$ or $\Gamma ^{\lambda }{}_{\mu \nu }=\frac{%
\partial x^{\prime \lambda }}{\partial x^{\tau }}\frac{\partial ^{2}x^{\tau }%
}{\partial x^{\prime \mu }\partial x^{\prime \nu }}$ must be introduced.

Just like any other interactions, gravity essentially has nothing to do with
the local Lorentz group $SO(1,3)$.\textbf{\ }Its localization amounts to an
accelerating transformation, the corresponding connection amounts to an
inertial force rather than a potential. Because of this, the local Lorentz
group cannot be considered as a gauge group in the sense of Yang-Mills
theories. All it does is restore the covariance of the theory. The
covariantisation of the teleparallel gravity is just a way to deal with
issues of gravity in a non-inertial reference system and then does not make
any physical change in the theory.

The only link between gravity and the Lorentz group is the equivalence
principle, which is due to the universality of the connection as an inertial
force. Both gravity and inertial have one common character, universality
which is absent of other fundamental interactions. The root of the
equivalence principle lies in universality of gravity. The root of the
universality lies in the Noether theorem, according to which the
energy-momentum is the current and source of gravity. It is the universality
of the energy-momentum that leads to the equivalence principle and the
difference between gravity and other interactions.

A pure gravitational theory ought to rid of inertial force. Neither
Einstein's Lagrangian nor the teleparallel Lagrangian include the curvature.
The only exception is the Hilbert Lagrangian which distinguish from
Einstein's Lagrangian by only a boundary term. In fact, in all the three
cases of\ the Geometrical Trinity of Gravity, nether the affine connection
nor the Lorentz connection is independent but a function of the metric or
the tetrad and their derivatives as remarked in 3.2.3. The fundamental
variable describing gravity is the translation connection $B^{I}{}_{\mu }$,
its field strength is the derivative $\partial _{\nu }B^{I}{}_{\mu }$. The
geometrization of gravity requires it to be expressed by a geometric object
including $B^{I}{}_{\mu }$ with well transformation character. Einstein's
genius is choosing the metric $g_{\mu \nu }$ in Riemann geometry and the
tetrad $e^{I}{}_{\mu }$ in Cartan geometry without knowing $B^{I}{}_{\mu }$.
Both of them are covariant under coordinate transformations as well as
Lorentz transformations. Einstein also constructed two simple and reasonable
Lagrangian (74) and (98), founded GR and initiated teleparallel gravity.

As a pure gravitational effect the expansion of the universe i.e. the
receding of galaxies and the fall of bodies from Galileo's Tower should be
caused by the same kind of force. Nevertheless, the force causing the fall
of bodies is sourced by the energy-momentum of our planet, while the force
causing the receding of galaxies is sourced by the energy-momentum of the
gravitational field itself. This is the so called geometry dark energy. In
the equations

\begin{equation*}
\partial _{\alpha }\left( \sqrt{-g}\Pi _{\lambda }^{\alpha \mu \nu \rho
\sigma }\left( g\right) L^{\lambda }{}_{\rho \sigma }\right) =\sqrt{-g}%
\left( t^{\mu \nu }+T^{\mu \nu }\right) ,\ \text{\ \ \ \ \ \ \ \ \ \ \ \ \ \
\ \ \ \ \ \ \ \ \ \ \ \ \ }\ \ \ (80)
\end{equation*}%
and%
\begin{equation*}
\partial _{\sigma }\left( eS_{I}{}^{\sigma \rho }\right) =\frac{1}{2}e\left(
t_{I}{}^{\rho }+\Theta _{I}{}^{\rho }\right) ,\text{ \ \ \ \ \ \ \ \ \ \ \ \
\ \ \ \ \ \ \ \ \ \ \ \ \ \ \ \ \ \ \ \ \ \ \ \ \ \ \ \ \ \ \ \ \ \ \ \ }\ \
\ (89)
\end{equation*}%
\ $t^{\mu \nu }$ and $t_{I}{}^{\rho }$ can be considered as the \textbf{%
geometry dark energy}. It is this \textbf{geometry dark energy that }\textsl{%
causes the accelerated expansion of the universe\ in }$f(R)$\textsl{, }$f(T)$%
\textsl{, }$f(Q)$\textsl{\ theories} [86-90].

\textbf{\ Acknowledgments }
 The research work is supported by the National Natural Science Foundation of China (12175095,12075109 and 11865012), and supported by  LiaoNing Revitalization Talents Program (XLYC2007047).

\textbf{\LARGE References}

[1] J. B. Jimenez, L. Heisenberg, and T. S. Koivisto, The Geometrical
Trinity of Gravity, Universe 5, 173 (2019), arXiv:1903.06830 [hep-th].

[2] R. Aldrovandi and J. G. Pereira, Teleparallel Gravity: An Introduction
(Springer, Dordrecht, The Netherlands, 2013)
https://doi.org/10.1007/978-94-007-5143-9.

[3] J. W. Maluf, \textquotedblleft The teleparallel equivalent of general
relativity,\textquotedblright\ Annalen Phys. (Berlin) 525, 339 (2013),
arXiv:1303.3897 [gr-qc].

[4] J. M. Nester and H. J. Yo, Symmetric teleparallel general relativity,
Chin. J. Phys. 37, 113 (1999), arXiv:gr-qc/9809049.

[5] J. Beltran Jimenez, L. Heisenberg, and T. Koivisto, Coincident General
Relativity, Phys. Rev. D 98, 044048 (2018), arXiv:1710.03116\ [gr-qc].

[6] S. Kobayashi and K. Nomizu, Foundations of Differential Geometry, 2nd
edition, Wiley--Intersciense, New York, 1996.

[7] M. Hohmann, Variational Principles in Teleparallel Gravity Theories,
Universe 7, 114 (2021), arXiv:2104.00536 [gr-qc].

[8] M. Adak, T. Dereli, T. S. Koivisto, and C. Pala, General teleparallel
metrical geometries, arXiv:2303.17812 [gr-qc].

[9] T. Koivisto, On an integrable geometrical foundation of gravity, Int. J.
Geom. Meth. Mod. Phys. 15, 1840006 (2018), arXiv:1802.00650 [gr-qc].

[10] T. Koivisto, M. Hohmann, and T. Z losnik, The General Linear Cartan
Khronon, Universe 5, 153 (2019), arXiv:1905.02967[gr-qc].

[11] F. W. Hehl, Four Lectures on Poincare Gauge Field Theory,
arXiv:2303.05366 [gr-qc].

[12] Y. Itin, F. W. Hehl and Y. N. Obukhov, Premetric equivalent of general
relativity: Teleparallelism, Phys. Rev. D 95, 084020 (2017),
arXiv:1611.05759 [gr-qc].

[13] T. Koivisto,\ Cosmology in the Lorentz gauge theory, arXiv:2306.00693\
[gr-qc].

[14] M. Blagojevic, F. W. Hehl, and T. W. B. Kibble, Gauge Theories of
Gravitation, IMPERIAL COLLEGE PRESS, 2013.

[15] T. Koivisto and T. Zlosnik, Paths to gravitation via the gauging of
parameterized field theories, arXiv:2212.04562\ [gr-qc].

[16] M. Kr\v{s}\v{s}\'{a}k, E.N. Saridakis. The covariant formulation of $%
f(T)$\ gravity. Classical and Quantum Gravity 33 115009 (2016),
arXiv:1510.08432 [gr-qc].

[17] A. Golovnev, T. Koivisto, M. Sandstad. On the covariance of
teleparallel gravity theories. Classical and Quantum Gravity 34, 145013
(2017), arXiv:1701.06271[gr-qc].

[18] J.W. Maluf, S.C. Ulhoa, J.F. da Rocha-Neto, F.L. Carneiro. Difficulties
of teleparallel theories of gravity with local Lorentz symmetry. Classical
and Quantum Gravity 37, 067003 (2020), arXiv:1811.06876 [gr-qc].

[19] C. Bejarano, R. Ferraro, F. Fiorini, M.J. Guzman. Reflections on the
covariance of modified teleparalleltheories of gravity. Universe 5 158 (2019)%
\U{ff0c}arXiv:1905.09913 [gr-qc].

[20] M. Kr\v{s}\v{s}\'{a}k, R.J. van den Hoogen, J.G. Pereira, C.G. Bohmer,
A.A. Coley. Teleparallel theories of gravity: illuminating a fully invariant
approach. Classical and Quantum Gravity 36 183001 (2019), arXiv:1810.12932
[gr-qc].

[21] A. Golovnev, The geometrical meaning of the Weitzenbock connection,
arXiv:2302.13599[gr-qc].

[22] M. Kr\v{s}\v{s}\'{a}k, Teleparallel Gravity, Covariance and Their
Geometrical Meaning, arXiv:2401.08106[gr-qc].

[23] R. Utiyama, Invariant theoretical interpretation of interaction. Phys.
Rev. 101, 1597 (1956).

[24] S. Weinberg, The Quantum theory of fields. Vol. 1: Foundations
(Cambridge University Press, 2005).

[25] T. Koivisto, On an integrable geometrical foundation of gravity,
arXiv:1802.00650v1[gr-qc].

[26] T. Koivisto, On an integrable geometrical foundation of gravity,
arXiv:1802.00650 [gr-qc].

[27] J. B. Jimenez, T. S. Koivisto, Lost in translation: the Abelian affine
connection (in the coincident gauge), arXiv:2202.01701\ [gr-qc].

[28] H. Jennen and J. G. Pereira, Dark energy as a kinematic effect,\ Phys.
Dark Univ., 11, 49 (2016)\ \ arXiv:1506.02012 [gr-qc].

[29] J. G. Pereira, Yuri N. Obukhov, Gauge Structure of Teleparallel
Gravity, Universe 5, 139 (2019), arXiv:1906.06287\ [gr-qc].

[30]. R. Aldrovadi and J. G. Preira, An Introduction to Geometrical Physics,
second edition; World Scientific, Singapore, 2017.

[31]. S. W. Hawking and G. F. R. Ellis, The Large Scale Structure of
Space-Time; Cambridge University Press, Cambridge, 1973.

[32] A. Marsh, Mathematics for Physics: An Illustrated Handbook; World
Scientific, Singapore, 2018.

[33] H. I. Arcos. and J. G. Pereira, TORSION GRAVITY: A REAPPRAISAL,
arXiv:gr-qc/0501017.

[34] A. Einstein, \textquotedblleft Hamiltonsches Prinzip und allgemeine
Relativit\"{a}tstheorie,\textquotedblright\ Sitzungsberichte der K\"{o}%
niglich Preu$\beta $ischen Akademie der Wissenschaften (Berlin), pp.
1111-1116 (1916).

[35] F. W. Hehl, J. D. McCrea, E. W. Mielke, and Y. Ne'eman, Metric affine
gauge theory of gravity: Field equations, Noether identities, world spinors,
and breaking of dilation invariance, Phys. Rept. 258, 1 (1995),
arXiv:gr-qc/9402012 [gr-qc].

[36] J. B. Jimenez, L. Heisenberg, and T. S. Koivisto, The canonical frame
of purified gravity, Int. J. Mod. Phys. D 28, 1944012 (2019),
arXiv:1903.12072 [gr-qc].

[37] A. Einstein, Riemann-Geometrie mit Aufrechterhaltung des Begriffes des
Fernparallelismus Sitzber. Preuss. Akad. Wiss. 17 217 (1928) .

[38] R. Aldrovandi, H. I. Arcos, and J. G. Pereira, General relativity as a
genuine connection theory, arXiv:gr-qc/0412032.

[39] M. Calcada\ and J. G. Pereira, Gravitation and the Local Symmetry Group
of Space-Time, arXiv:gr-qc/0201059.

[40] T. Koivisto, and T. Zlosnik, Paths to gravitation via the gauging of
parameterized field theories, arXiv:2212.04562 [gr-qc]

[41] Jia-An Lu, On the dierence between Poincare and Lorentz gravity, Gen.
Rel. Grav. 49, 138 (2017), arXiv:1705.00297 [gr-qc].

[42]\ A. Einstein, The Meaning of Relativity, Princeton Lectures of May
1921, 5th ed. (Princeton Univ. Press, Princeton, NJ, 1955.

[43] A. Einstein, Riemann-Geometrie mit Aufrechterhaltung des Begriffes des
Fernparallelismus, Sitzber. Preuss. Akad. Wiss. 17 (1928) 217--221.

[44] A. Einstein, Neue M\"{o}glichkeit f\"{u}r eine einheitliche Feldtheorie
von Gravitation und Elektrizit\"{a}t, Sitzber. Preuss. Akad. Wiss. 17 (1928)
224--227.

[45] A. Einstein, Einheitliche Feldtheorie und Hamiltonsches Prinzip ,
Sitzber. Preuss. Akad. Wiss. 18 156 (1929).

[46] Weitzenb\"{o}ck, Roland, Differentialinvarianten in der Einsteinschen
Theorie des Fernparallelismus, Sitzungsber. Preuss. Akad. Wiss. Berlin 26
466 (1916).

[47] R. Weitzenb\"{o}ck, Invariantentheorie. P. Noordhoff, Groningen, 1923.

[48] S. Bahamonde, K. F. Dialektopoulos, C. Escamilla-Rivera, G. Farrugia,
V. Gakis, M. Hendry, M. Hohmann, J. Levi Said, J. Mifsud, and E. Di
Valentino, Teleparallel Gravity: From Theory to Cosmology, arXiv:2106.13793
[gr-qc].

[49] J. Boos, F. W. Hehl, Gravity-induced four-fermion contact interaction
implies gravitational intermediate W and Z type gauge bosons. Int. J. Theor.
Phys. 56, 751 (2017), https://doi.org/10.1007/s10773-016-3216-3.

[50] T. W. B. Kibble, Lorentz invariance and the gravitational field. J.
Math. Phys. 2, 212 (1961) https://doi.org/10.1063/1.1703702.

[51] F. W. Hehl, P. von der Heyde, G. D. Kerlick, and J. M. Nester, Rev.
Mod. Phys. 48:393 (1976).\ \ \ 

[52] F. W. Hehl and Y. N. Obukhov, Conservation of energy-momentum of matter
as the basis for the gauge theory of gravitation, arXiv:1909.01791 [gr-qc].

[53] H. I. Arcos, T. Gribl Lucas and J. G. Pereira, Consistent
Gravitationally-Coupled Spin-2 Field Theory, Class. Quantum Grav. 27, 145007
(2010), arXiv:1001.3407 [gr-qc].

[54] J. J Sakurai, Theory of strong interactions. Ann. Phys. (N.Y.) 11, 1
(1960) https://doi.org/10.1016/0003-4916(60)90126-3.

[55] S. L. Glashow, M. Gell-Mann, Gauge theories of vector particles. Ann.
Phys. (USA) 15, 296-297 (1961) https://doi.org/10.1016/0003-4916(61)90020-3.

[56] R. Feynman, F. B. Morinigo,W. G.Wagner, Feynman Lectures on
Gravitation, Lectures given 1962/63, B. Hatfield, ed. (Addison-Wesley,
Reading, MA, 1995)

[57] K. Hayashi and T. Nakano, Extended translation invariance and
associated gauge fields, Prog. Theor. Phys. 38 491--507 (1967).

[58] Y. M. Cho, Einstein Lagrangian as the translational Yang-Mills
Lagrangian, Phys. Rev. D 14, 2521 (1976)
https://doi.org/10.1103/PhysRevD.14.2521.

[59] J. Nitsch, F. W. Hehl, Translational gauge theory of gravity:
Postnewtonian approximation and spin precession. Phys. Lett. B 90, 98 (1980)
https://doi.org/10.1016/0370-2693(80)90059-3.

[60] A. Golovnev, Introduction to teleparallel gravities, Proceedings of the
9th Mathematical Physics Meeting: School and Conference,
arXiv:1801.06929[gr-qc].

[61] F. W. Hehl and Y. N. Obukhov, Conservation of energy-momentum of matter
as the basis for the gauge theory of gravitation, arXiv:1909.01791 [gr-qc].

[62] H. Jennen and J. G. Pereira, Dark energy as a kinematic effect,\ \
arXiv:1506.02012 [gr-qc].

[63] T. Koivisto, Energy in the Relativistic Theory of Gravity, arXiv:\
2202.11522.

[64] B. Li, T. P. Sotiriou, and J. D. Barrow, $f(T)$\ gravity and local
Lorentz invariance, Phys. Rev. D83 064035 (2011) , arXiv:1010.1041 [gr-qc].

[65] M. Blagojevi\'{c} and J. M. Nester, From the Lorentz invariant to the
coframe form of $f(T)$ gravity, arXiv:2312.14603 [gr-qc].

[66] D. Blixt, A. Golovnev, M.J. Guzman, R. Maksyutov, Geometry and
covariance of symmetric teleparallel approaches to gravity, arXiv:2306.09289
[gr-qc].

[67] A. Golovnev, Issues of Lorentz-invariance in $f(T)$\ gravity and
calculations for spherically symmetric solutions, arXiv:2105.08586 [gr-qc].

[68] D. Blixt, R. Ferraro, A. Golovnev and M. J. Guzman, Lorentz
gauge-invariant variables in torsion-based theories of gravity, Phys. Rev. D
105, 084029 (2022), arXiv:2201.11102 [gr-qc].

[69] A. Golovnev, M. J. Guzman, Lorentz symmetries and primary constraints
in covariant teleparallel gravity, Phys. Rev. D 104, 124074 (2021),
arXiv:2110.11273 [gr-qc].

[70] M. Hohmann, L. Jarv, M. Kr\v{s}\v{s}\'{a}k and C. Pfeifer, Teleparallel
theories of gravity as analogue of non-linear electrodynamics, Phys. Rev. D
97, 104042 (2018), arXiv: 1711.09930 [gr-qc].

[71] M. Hohmann, L. Jarv and U. Ualikhanova, Covariant formulation of
scalar-torsion gravity, Phys. Rev. D 97, 104011 (2018), arXiv: 1801.05786
[gr-qc].

[72] A. Golovnev and M. J. Guzman, Foundational issues in $f(T)$ gravity
theory, Int. J. Geom. Meth. Mod. Phys. 18 (2021) no.supp 01, 2140007
[arXiv:2012.14408].

[73] C. G. Boehmer and E. Jensko, Modified gravity: A unified approach,
Phys. Rev. D 104, 024010 (2021), arXiv:2103.15906 [gr-qc].

[74] C. G. Boehmer and E. Jensko, Modified gravity: a unified approach to
metric-affine models, arXiv:2301.11051 [gr-qc].

[75] E. Kretschmann, Uber den Physikalischen Sinn der Relativit tspostulate,
Annalen Der Physik 53 (1917) 575-614.

[76] Bondi, Hermann; in: Relativity, Quanta, and Cosmology, edited by: M.
Pantaleo and F. de Finis; vol. 1 p. 181; Johnson Reprint Corporation; New
York (1979)\ 

[77] D. A. Gomes, J. B. Jimenez and T. S. Koivisto, General Parallel
Cosmology, arXiv:2309.08554 [gr-qc].

[78] D.W. Sciama, The analogy between charge and spin in general relativity.
In: Recent Developments in General Relativity, Festschrift for Infeld
(Pergamon Press, Oxford; PWN, Warsaw, (1962) 415.

[79] D. Lynden-Bell, in Variable Stars and Galaxies, in honor of M. W. Feast
on his retirement, Astronomical Society of the Pacific Conference Series,
Vol. 30, edited by B. Warner (1992) p. 1.

[80] Z. Oziewicz, J. Phys. Conf. Ser. 532, 012021 (2014).

[81] J. B. Jimenez, T. S. Koivisto, Lost in translation: the Abelian affine
connection (in the coincident gauge), arXiv:2202.01701\ [gr-qc].

[82] J. D. Norton, Reports on Progress in Physics 56, 791 (1993).

[83] R. M. Wald, General Relativity, The Univercity of Chicago Press,
Chicago and London, 1984.

[84] A. Einstein, Uber das Relativitats Prinzip und die aus demselben
gezogenen Folgerungen [On the Relativity Principle and the Conclusions
drawnfrom it], Jahrbuch der Radioaktivitat und Elektronik, 4, pp. 411 (1907).

[85] J. L. Synge, "Relativity, The General Theory" North Holland Publish ing
Co, Amsterdam, The Netherlands (1960).

[86] S. Capozziello and M. Francaviglia, Extended Theories of Gravity and
their Cosmological and Astrophysical Applications,~Gen.Rel.Grav.40, 357
(2008), arXiv:0706.1146~[astro-ph].

[87] L. G. Jaime, L. Patino, M. Salgado, Note on the equation of state of
geometric dark-energy in $f(R)$ gravity, Phys. Rev. D.89 084010 (2013),
arXiv:1312.5428 [gr-qc].

[88] J. Lu and G. Chee, Cosmology in Poincare gauge gravity with a
pseudoscalar torsion, JHEP 05 024 (2016), arXiv:1601.03943 [gr-qc].

[89] E. V. Linder, Einstein's Other Gravity and the Acceleration of the
Universe, Phys. Rev. D81 127301 (2010), arXiv:1005.3039 [astro-ph.CO].

[90] J. Lu, X. Zhao and G, Chee, Cosmology in symmetric teleparallel gravity
and its dynamical system, Eur. Phys. J. C 79, 530 (2019), arXiv:1906.08920
[gr-qc].

\end{document}